\documentstyle[12pt,aaspp4]{article}
\begin{document}
\title{RR Lyrae stars in Galactic globular clusters. II.
A theoretical approach to variables in M3}
\author{Marconi, M.\altaffilmark{1}, Caputo, F.\altaffilmark{2},
Di Criscienzo, M.\altaffilmark{1}, and Castellani,
M.\altaffilmark{2}} \altaffiltext{1}{INAF - Osservatorio
Astronomico di Capodimonte, via Moiarello 16, I-80131 Napoli,
Italy; marcella@na.astro.it, dicrisci@na.astro.it}
\altaffiltext{2}{INAF - Osservatorio Astronomico di Roma, Via di
Frascati, 33, I-00040 Monte Porzio Catone Italy;
caputo,mkast@coma.mporzio.astro.it}
\begin{abstract}

We present predicted relations connecting pulsational (period and
amplitude of pulsation) and evolutionary (mass, absolute magnitude
and color) parameters, as based on a wide set of convective
pulsating models of RR Lyrae stars with $Z$=0.001, $Y$=0.24 and
mass and luminosity suitable for the ``old'' (age $>$ 8 Gyr)
variables observed in globular clusters. The relations are
collated with sound constraints on the mass of pulsators, as
inferred from up-to-date evolutionary models of horizontal branch
stars, in order to provide a self-consistent theoretical framework
for the analysis of observed variables. The theoretical
predictions are tested through a detailed comparison with
measurements of RR Lyrae stars in the globular cluster M3. We show
that the predicted relations satisfy a variety of observed data,
thus providing a $pulsational$ route to the determination of
accurate distances to RR Lyrae-rich globular clusters with
intermediate metal content. We show that current uncertainties on
the intrinsic luminosity of up-to-date  horizontal branch models,
as due to the input physics used in the computations by the
different authors, have a quite low influence ($\sim$ 0.02-0.03
mag) on the pulsational distance modulus. On the contrary, the
effect of the different bolometric corrections adopted to convert
bolometric luminosity into absolute magnitude is of the order of
$\sim$ 0.05-0.06 mag. The pulsation models are also used to
perform some valuable tests on the physics adopted in current
stellar evolution computations. We show that the constraints
inferred by pulsation theory support the large value of the
mixing-length parameter ($l/H_p$=1.9-2.0) adopted to fit observed
Red Giant Branches, but, at the same time, they would yield that
the luminosity of the horizontal branch updated models is too
bright by $\sim$ 0.08$\pm$0.05 mag, if helium and metal diffusion
is neglected. Conversely, if element diffusion is properly taken
into account, then  there is a marginal discrepancy of $\sim$
0.04$\pm$0.05 mag between evolutionary and pulsational
predictions.
\end{abstract}
\section{Introduction}
Radial pulsating variables, whether they are Population I or
Population II stars, are of great relevance in several fields of
modern astrophysics. In particular, RR Lyrae variables are widely
used, via the calibration of their absolute visual magnitude
$M_V$(RR) in terms of the iron-to-hydrogen content [Fe/H], as
standard candles for distance determination in the Local Group. On
this basis, they provide an independent test of the Cepheid
distance scale for nearby galaxies (Magellanic Clouds, M31) as
well as a calibration of secondary distance indicators (e.g., the
globular cluster luminosity function) in external galaxies, thus
yielding relevant clues about the value of the Hubble constant. On
the other hand, since the absolute magnitude of the globular
cluster main sequence turn-off  is a classical ``clock'' to estimate
the age of these ancient stellar systems, one easily understands
the relevance of an accurate RR Lyrae distance scale for
cosmological studies.

Accordingly, a huge amount of work has been made in the last years
to establish the absolute magnitude of Horizontal Branch (HB)
stars, and particularly of RR Lyrae variables, and its dependence
on the star metal content. Unfortunately, a general consensus has
not been achieved yet, leading to the well-known debate between
the so-named ``short'' and ``long'' distance scale.

On the observational side (e.g., Baade-Wesselink method, Hipparcos
and statistical parallax), the estimates of $M_V$(RR) suggest
relatively faint magnitudes, although leaving significant
uncertainties ($>$ 0.1 mag, see Vandenberg et al. 2000, hereafter
VDB, and references therein). As a matter of example, in the
recent review by Cacciari (1999) the empirical values of $M_V$(RR)
at [Fe/H]=$-$1.6 actually range from $\sim$ 0.4 to $\sim$ 0.7 mag.
Moreover, it should be mentioned that Fernley et al. (1998) and
Groenewegen \& Salaris (1999), by assuming the same slope $\Delta
M_V$(RR)/$\Delta$ [Fe/H]=0.18 mag/dex, give respectively
$M_V$(RR)=0.78$\pm$0.15 mag and 0.50$\pm$0.26 mag at
[Fe/H]=$-$1.5, as derived from an identical sample of variables
but using statistical parallaxes or reduced parallaxes,
respectively. In reality, not only the zero-point but also the
slope of the $M_V$(RR)-[Fe/H] relation is matter of considerable
debate and in the relevant literature one finds values in the
range of $\sim$ 0.13 to $\sim$ 0.30 mag/dex.

As for the theoretical studies based on HB models, all recent
computations that adopt up-to-date physics suggest absolute
magnitudes which are somehow brighter than empirical estimates, at
fixed [Fe/H]. However, evolutionary predictions themselves, as
provided by the various researchers, still suggest significant
discrepant (both in the zero-point and the slope) 
$M_V$(RR)-[Fe/H] relations (see VDB; Caputo et al. 2000
and references therein). On this issue, it should be noted that
HB evolutionary models provide a
relation between the bolometric luminosity $L$ and the overall metallicity
$Z$. Thus, any comparison among the various evolutionary predicted 
$M_V$(RR)-[Fe/H] relations 
is affected by two factors of
critical importance: a) the conversion of bolometric luminosity
and effective temperature into magnitude and color, i.e. the
adopted bolometric correction and temperature-color
transformation, and b) the scaling between the theoretical $Z$ and
the measured [Fe/H] value, i.e. the adopted ratio between
$\alpha$-elements and iron. These factors are still more important
when theoretical predictions are compared with observed
quantities. In addition, we wish also to mention that the constraints inferred 
by evolutionary
models  refer to ``$static$'' stars, whereas the RR Lyrae
observed magnitudes are ``$mean$'' quantities averaged over the
pulsation cycle, which not necessarily are exactly alike static
values.

In this context, it is of great help to consider simultaneously
the double nature of RR Lyrae stars - from one side, radial
pulsators, from the other one, low-mass ($M\le$
0.85$M_{\odot}$) He-burning stars - since the pulsational
quantities (i.e. period and amplitude of pulsation) depend on the
star structural parameters (mass, luminosity and effective
temperature),  and are firm and safe observables as well, fully
independent of distance and reddening.

On this ground, we planned to construct an exhaustive and
homogeneous theoretical framework, both on the pulsational and the
evolutionary side, aiming at a sound analysis of observed RR Lyrae
stars at various metal contents (see Castellani, Caputo \&
Castellani 2003, Paper I, for the observational scenario of
variables in Galactic globular clusters). Extensive sets of
nonlinear convective pulsation models (Di Criscienzo, Marconi \&
Caputo 2003, Paper III, in preparation) and HB evolutionary tracks
(Cassisi, Castellani, Caputo \& Castellani 2003, Paper IV, in
preparation) are computed using the same input physics. Then,
adopting similar bolometric corrections and color-temperature
transformations, self-consistent predictions concerning various
stellar parameters are derived.

The general procedure is discussed in the present paper, which
deals with the results at $Z$=0.001 and $Y$=0.24. We selected this
metal content as suitable for the analysis of several RR Lyrae
rich globular clusters, in particular M3 which is the most RR
Lyrae rich cluster in the Galaxy and among the best studied ones.
The organization of the paper is as follows: Section 2 presents
the pulsation models and gives the relevant relations among the
various parameters, while the constraints from stellar evolution
theory are discussed in Section 3. The comparison with available $BVK$
photometry 
of RR Lyrae variables in M3 is presented in Section 4. Discussion
and some brief remarks close the paper.

\section{Theoretical pulsational scenario}

\subsection{Pulsation models}

The nonlinear convective hydrodynamical code used for pulsation
model computations has been already described in a series of
papers (see, e.g., Bono \& Stellingwerf 1994; Bono et al. 1997a,b)
and it will  not be discussed further. We wish only to mention that
the models adopt recent opacity compilations by Iglesias \& Rogers
(1996) for temperatures higher than 10,000 K and by Alexander \&
Ferguson (1994) for lower temperatures. As in previous papers, a
mixing length parameter $l/H_p$=1.5 is adopted in order to close
the system of convective and dynamic equations. However, to gain
some insight into how the boundaries for instability are dependent
on convection (see Caputo et al. 2000), additional models with
$l/H_p$=2.0 are computed, also in view of the values for the
mixing length parameter recently used in HB model computations
(see Sect. 3).

For the pulsation models with $Z$=0.001, we adopt two masses
($M$=0.65 and 0.75$M_\odot$) and three luminosity levels
(log$L/L_\odot$=1.51, 1.61, 1.72) in order to explore the
dependence on the stellar mass and luminosity. For each given set
of entry parameters, the computations are performed by decreasing
the effective temperature $T_e$ by steps of 100-200 K. In the
following, we refer to the results from the first (bluest)
pulsating model to the last (reddest) one. Let us be clearly
understood that increasing (decreasing) by 100 K the effective
temperature of the bluest (reddest) pulsators yields non-pulsating
structures.

Since the adopted nonlinear approach supplies not only the
pulsation period, but also the luminosity variation along the
whole pulsation cycle, the bolometric light curves are transformed
into the observational plane by adopting bolometric corrections
and temperature-color transformations  provided by Castelli,
Gratton \& Kurucz (1997a,b, hereafter CGKa,b). In this way,
amplitudes and $mean$ absolute magnitudes, either
intensity-weighted and magnitude-weighted, are derived in the
various photometric bands.

Table 1 lists selected results\footnote{The entire set of models
can be obtained upon request to
marcella@na.astro.it} for fundamental (F) pulsators with
0.65$M_\odot$, log$L/L_\odot$=1.61 and $l/H_p$=1.5. For each given
effective temperature, we give the static magnitude, i.e. the
value the star would have were it not pulsating, the period $P$
(in days), mean intensity-weighted $<M_i>$ and magnitude-weighted
($M_i$) magnitudes, and amplitudes $A_i$. Table 2 has the same
meaning, but holds for first-overtone (FO) pulsators.

As well known since the original Ritter's formulation
$P\rho^{1/2}$=$Q$ ($\rho$ is the star density and $Q$ the
pulsation constant), the basic physics underlying radial
variability suggests that the pulsation period depends on the
mass, luminosity and effective temperature of pulsators. Linear
interpolation among all our data supplies a refined version of
earlier pulsation equations (see, e.g.,  van Albada \& Baker 1971,
Bono et al. 1997a), as given by

$$\log P^F=11.066(\pm 0.002)+0.832\log L-0.650\log M-3.363\log T_e\eqno(1a)$$
$$\log P^{FO}=10.673(\pm 0.001)+0.805\log L-0.603\log M-3.281\log T_e\eqno(1b)$$

\noindent where luminosity $L$ and mass $M$ are in solar units.
According to our models, the period of FO pulsators can be
``fundamentalized'' as log$P^F-$log$P^{FO}$ $\sim$ 0.13, on average,
or with the more punctual relation
log$P^F$=0.147+1.033log$P^{FO}$, neglecting the weak dependence on
mass and luminosity which plays a significant role only when
dealing with double mode pulsators.

The natural outcome of the pulsation equation in the observational
plane is the  Period-Magnitude-Color ($PMC$) relation in which the
pulsation period, for any given mass, is correlated with the
pulsator absolute magnitude and color. Using $BV$ static
quantities\footnote{We give here only the predicted relations which are 
used in comparison with $BVK$ data of M3 RR Lyrae stars. As for the 
relations in other photometric bands, they can be requested to 
marcella@na.astro.it.}, 
linear interpolation through the results (see Fig. 1) gives

$$\log P^F=-0.612(\pm 0.002)-0.340M_V-0.663\log M+1.307[M_B-M_V]\eqno(2a)$$
$$\log P^{FO}=-0.762(\pm 0.001)-0.340M_V-0.643\log M+1.381[M_B-M_V],\eqno(2b)$$

\noindent

According to the predicted $PMC$ relations, for a sample of RR
Lyrae stars at the same distance and with the same reddening, e.g.
for variables in a given globular cluster with no differential
reddening, one could estimate the mass range spanned by the
variables with an uncertainty lower than 2\%, once periods and
static colors are firmly known. If the cluster distance modulus
and reddening are independently determined, then the mass absolute
values can be inferred too. Alternatively, once period and color
are measured, the relations can be used to get the distance to
individual RR Lyrae stars with known mass and reddening.

Let us now consider the quite plain correlation between the
absolute magnitude in a given $i$-band and luminosity,
$M_i=-$2.5log$L$+$BC_i$, where the bolometric correction $BC_i$
depends on the star effective temperature, mass and luminosity.
Since also the pulsation period is governed [see eqs. (1a) and
(1b)] by the effective temperature, mass and luminosity, the
absolute magnitude $M_i$ of radial pulsating stars is expected to
be a function of mass, luminosity and period. It is obvious that
the dependence of $M_i$ on mass, luminosity and period is governed by
the dependence of $BC_i$ on the effective temperature.

We show in Fig. 2 (symbols as in Fig. 1) the  quite different
dependence of visual and near-infrared static absolute magnitudes
on the pulsation period. For any given mass and luminosity, an
increase of $T_e$ yields shorter periods, while $M_V$ becomes
slowly brighter and $M_K$ significantly fainter. On the other
hand, if increasing the luminosity at constant mass and effective
temperature, the period increases and both $M_V$ and $M_K$ become
brighter. In summary, one has a sort of $degeneracy$ (see the
arrows in the lower panel in Fig. 2)  between the luminosity and
effective temperature effects on $M_K$. As a result, for fixed
period, $M_V$ depends mainly on luminosity, with a negligible
dependence on the mass, whereas $M_K$ shows small but non
negligible variations both with luminosity and mass (see also Bono
et al. 2001, 2003). According to present models, the predicted
period-magnitude relations in the visual ($PM_V$) and
near-infrared bandpass ($PM_K$) are

$$M_V^F= 5.597(\pm 0.002)+0.415\log P^F+0.305\log M-2.890\log L\eqno(3a)$$
$$M_V^{FO}=5.478(\pm 0.002)+0.319\log P^{FO}+0.253\log M-2.818\log L\eqno(3b)$$

\noindent and

$$M_K^F=-0.182(\pm 0.002)-2.197\log P^F-1.447\log M-0.663\log L\eqno(4a)$$
$$M_K^{FO}=-0.502(\pm 0.002)-2.272\log P^{FO}-1.379\log M-0.659\log L\eqno(4b)$$
The importance of the RR Lyrae near-infrared $PM_K$ relation to
get accurate distances to globular clusters has already been
discussed by several authors (see Bono et al. 2001 and references
therein) and it will not be discussed further. It is enough to
mention that near-infrared magnitudes are very little affected by
reddening and that adopting for the RR Lyrae stars in a given
globular cluster, that is for variables at constant age and metal
content, a luminosity dispersion of $\delta$ log$L\sim \pm$ 0.06
and a mass spread\footnote{Based on current computations, the mass
spread of HB models near the RR Lyrae instability strip depends on
$Z$, decreasing with increasing the metal content.} of $\pm$3\%,
yields a dispersion $\delta M_V \sim \pm$ 0.17 mag and $\delta M_K
\sim \pm$ 0.05 mag, at fixed period.

\subsection{The instability strip}

All the relations described in the previous section originate from
the pulsation equation which formally fixes the pulsation period for
any given combination of the entry parameters ($M$, $L$, $T_e$).
However, it is well known that radial pulsation occurs in a quite
well-defined region of the HR diagram, the so-named ``instability
strip'', that depends on the type of pulsation and the intrinsic
parameters of pulsators. In simple words, pulsation does not occur
for any combination of $M, L$ and $T_e$. For each given mass and
luminosity there is a maximum and minimum effective temperature
for the onset of either fundamental and first-overtone pulsation.

As already known, the data in Tables 1 and 2,
as those presented in previous papers
(see, e.g., Bono, Caputo \& Marconi 1995a, Bono et
al. 1997a,b, Caputo et al. 2000, Bono et al. 2002 and references
therein) show that, for fixed mass and luminosity, FO pulsators
are generally bluer than F pulsators, in full agreement with the
observed behavior of $c$-type and $ab$-type RR Lyrae stars. This
yields that the first-overtone blue edge (FOBE) and the
fundamental red edge (FRE) can be taken as limits of the whole
instability strip. Moreover, the models show that the fundamental
pulsation blue edge (FBE) is generally bluer than the
first-overtone red edge (FORE), suggesting that both the pulsation
modes can be stable in a middle region of the instability region.
Such a result supports the hypothesis that the transition between
$ab$ and $c$ type RR Lyrae stars may occur along different
transitions (FBE or FORE), as well as that a hysteresis mechanism,
as earlier suggested by van Albada \& Baker (1973), could explain
the Oosterhoff dichotomy (see, e.g., Bono et al. 1995a, Bono et
al. 1997a for details).

Given the steps of 100 K in our computations, we adopt as edges of
the pulsation region the effective temperature of the bluest FO
model and of the reddest F model, increased and decreased by 50K,
respectively.   Then, linear interpolation to all the data with
$Z$=0.001 and $l/H_p$=1.5 gives the following relations

$$\log T_e^{FOBE}=3.970(\pm 0.004)-0.057\log L+0.094\log M\eqno(5)$$
$$\log T_e^{FRE}=3.957(\pm 0.007)-0.102\log L+0.073\log M\eqno(6)$$

\noindent where the uncertainties include the intrinsic
uncertainty of $\pm$50K on the FOBE and FRE temperatures.

Before proceeding, it is worth mentioning that a further source of
uncertainty on the pulsation limits is due to the efficiency of
convection in the star external layers. As the depth of convection
increases from high to low effective temperature, and the effect
of convection is to quench pulsation, we expect that varying the
value of the mixing length parameter will modify the red edge of
the instability strip by a larger amount with respect to the blue
edge. Model computations made with $l/H_p$=2.0 show that, at
constant mass and luminosity, the FOBE temperature decreases by
$\sim$ 100 K, whereas those at FBE and FRE increase by $\sim$ 100
K and $\sim$ 300 K, respectively. As a consequence, at $l/H_p$=2.0
the zero-point in eq. (5) and eq. (6) varies by $-$0.006 and
+0.020, respectively. Note that the knowledge of FOBE and FRE
effective temperatures  is basic to select variable and non
variable stars in synthetic HB populations (see Sect. 3.2)

For comparison with observed RR Lyrae stars, of particular
interest (see Caputo 1997; Caputo et al. 2000) is the location of
the pulsation theoretical edges  in the magnitude-period and
color-period diagrams. From present models at $Z$=0.001 and
$l/H_p$=1.5, and adopting CGKa,b atmosphere models, we derive

$$M_V^{FOBE}=-1.193(\pm 0.034)-2.230\log M-2.536\log P^{FO}\eqno(7)$$
$$M_V^{FRE}=0.119(\pm 0.040)-1.883\log M-1.962\log P^F\eqno(8)$$

\noindent and

$$[M_B-M_V]^{FOBE}=0.25(\pm 0.02)-0.10\log M+0.10\log P^{FO}\eqno(9)$$
$$[M_B-M_V]^{FRE}=0.48(\pm 0.03)-0.06\log M+0.27\log P^F\eqno(10)$$

\noindent where the uncertainties include the intrinsic $\pm$50 K
uncertainty on the FOBE and FRE temperatures. As a consequence of
the above mentioned effects of a larger mixing-length parameter on
the FOBE and FRE effective temperatures, adopting $l/H_p$=2.0 the
zero-point in eq. (7) and eq. (8) varies by $\sim$ +0.05 mag and
$\sim -$0.14 mag, respectively, while in eq. (9) and eq. (10) by
$\sim$ +0.02 mag and $\sim -$0.06 mag, respectively.

\subsection{Pulsation amplitudes}

We have already mentioned that the non-linear approach supplies
important clues on a further observable: the pulsation amplitude.
A glance at the  data in Table 1 and Table 2 (see also Bono et al.
1997a) shows that, for fixed mass and luminosity, there is a
direct correlation between fundamental amplitude and effective
temperature: the amplitude increases when moving from the red to
the blue limit for fundamental pulsation, that is as the period
decreases. As for FO pulsators, they show a
characteristic ``bell'' shaped distribution,
with the maximum amplitude attained in the middle of the
first-overtone region. As a whole, these theoretical features
appear consistent with observed RR Lyrae period-amplitude
diagrams, usually named ``Bailey diagrams'' after their inventor
(Bailey 1899, 1902).

Taking advantage of the quite linear correlation between
fundamental amplitude and period (logarithmic scale), we show in
the left panel of Fig. 3 the $A_V$-log$P$ distribution of F
pulsators at $l/H_p$=1.5 and different mass and luminosity
(symbols as in Fig. 1). One has that, for fixed period, the
amplitude decreases as the luminosity decreases or, by a rather
small amount, as the mass increases. As a whole, we show in the
right panel in Fig. 3 that all the results with $l/H_p$=1.5
(filled circles) can be approximated with the linear relation
(solid line)

$$\log P^{F*}=\log P^F+0.385M_V+0.299\log M=0.132(\pm0.02)-0.189A_V\eqno(11)$$

However, we wish to remark that the fundamental pulsation amplitude is not
``$governed$'' by the absolute value of the effective temperature,
but it depends on the ``$relative$'' location of the pulsator with
respect to the red limit (where vanishing amplitudes are attained)
and blue limit (where the amplitude reaches it maximum value) of
fundamental pulsation (see also Bono et al. 1997a).  As a matter of
example, the visual amplitude of the 0.65$M_{\odot}$ pulsator with
$T_e$=6500 K increases from $A_V$=0.77 to 1.05 mag as the
luminosity increases from log$L/L_{\odot}$=1.51 to 1.72. It
follows that the above described effects of a larger mixing-length
parameter on the effective temperature of the fundamental
pulsation limits will modify both the zero-point and the slope of
the predicted Period-Magnitude-Amplitude ($PMA$) relation. As a
fact, the right panel in Fig. 3 shows that the results with
$l/H_p$=2.0 (open circles) can be approximated as (dotted line)

$$\log P^{F*}=\log P^F+0.385M_V+0.349\log M=0.024(\pm0.02)-0.142A_V\eqno(12)$$

Closing this section on the pulsation amplitude, we present in the
left panel in Fig. 4 the relation between visual amplitude $A_V$
and static $BV$ color for  F pulsators with $l/H_p$=1.5 (symbols
as in Fig. 1). One derives that, at least in the explored range of
mass and luminosity, all the results can be approximated with a
common Color-Amplitude ($CA_{BV}$) relation, as given by (see the
solid line in the right panel)

$$M_B-M_V=0.49(\pm0.02)-0.137A_V\eqno(13)$$

However, for the reasons above discussed, also the
$CA_{BV}$ relation turns out to depend on the adopted
mixing-length parameter, becoming

$$M_B-M_V=0.42(\pm0.02)-0.113A_V\eqno(14)$$

\noindent at $l/H_p$=2.0 (see the dotted line in the right panel).

According to the predicted $CA_{BV}$ relation, the intrinsic $BV$
color of individual pulsators can be estimated within $\pm$0.02
mag, which means a similar accuracy in the knowledge of the
reddening $E(B-V)$ and a determination of the visual extinction
correction A(V) within $\sim \pm$ 0.06 mag. This might be of
particular relevance in the case of RR Lyrae variables in globular
clusters, or stellar fields, with highly differential reddening.

\subsection{Static and mean magnitudes}

All the relations presented so far, although involving
observable quantities such as magnitude and color, cannot be
straightaway applied to observed RR Lyrae variables because they
hold for $static$ values whereas empirical measurements refer to
quantities averaged over the pulsation cycle.

From the data listed in Table 1 one finds that the synthetic mean
magnitudes ($M_i$) of fundamental pulsators, as derived averaging
the magnitude over the cycle, are fainter than the corresponding
static values with a discrepancy increasing as the pulsation
amplitude increases. 
The key element governing such a behavior is the overall
morphology of the light curve in the sense that the difference
between static and mean magnitudes increases from symmetric (low
amplitude) to asymmetric (high amplitude) light curves, namely
moving from the red to the blue fundamental edge (see also Bono,
Caputo, Stellingwerf 1995b). Since at fixed effective temperature
the amplitude decreases from blue to visual to infrared, the
difference between static and mean values for $K$ magnitudes is
almost negligible, whereas $(M_B)$, $(M_V)$ and $(M_I)$ differ
from the corresponding static values up to $\sim$ 0.18 mag, $\sim$
0.14 mag, and $\sim$ 0.08 mag, respectively, at $A_V$=1.6 mag. As
for FO pulsators (data in Table 2), the discrepancy is less
pronounced, given their low-amplitude and quite symmetric light
curves. Note that present synthetic magnitudes confirm previous
$BVK$ results inferred by Bono et al. (1995b) from a preliminary
set of pulsating models with $Z$=0.001, but adopting $Y$=0.30,
older opacity tables (Cox \& Tabor 1976)  and Kurucz (1992)
stellar atmosphere models.

Concerning the intensity-averaged magnitudes $<M_i>$, data in Table 1 and 
Table 2 confirm that they are quite close to the static values, except for 
models with very large amplitude.

In Fig. 5 we show the synthetic visual magnitudes for F and FO
pulsators, by plotting the discrepancy between static and mean
values, as well as the internal difference between the two means,
as a function of the visual amplitude $A_V$. It is quite relevant
to notice that the predicted differences $(M_V)-<M_V>$, as well as
the corresponding rather insignificant amounts inferred from Table
1 and Table 2 for the $K$-band, are in close agreement with
observed differences $(V)-<V>$ and $(K)-<K>$, as reported in
previous studies on RR Lyrae stars (e.g. Fernley 1993).

As for the mean colors $m_i-m_j$ which are usually estimated using
either $(m_i)-(m_j)$ or $<m_i>-<m_j>$, since most of empirical
work on RR Lyrae stars refers to $B$ and $V$ magnitude, we show in
Fig. 6 the predicted discrepancy between static and mean $B-V$
color, as well as the difference between the two means $(M_B-M_V)$
and $<M_B>-<M_V>$. At variance with the behavior shown in Fig. 5,
one has that the $(M_B-M_V)$ color is redder, whereas
$<M_B>-<M_V>$ is bluer, than the static color. The evidence for a
positive difference between predicted $(M_B-M_V)$ and
$<M_B>-<M_V>$, with an amount that increases with $A_V$, agrees
with quite well settled observed trends (see, e.g. Liu \& Janes
1990; Storm, Carney \& Beck 1991; Carney et al. 1992, Caputo et
al. 1999).

According to the data presented in Fig. 5 and Fig. 6, mean
magnitudes and colors can be corrected for the amplitude effect to
get static values (see also Bono et al. 1995b). However, given the
different behavior of $M_i-(M_i)$ and $M_i-<M_i>$ with the
amplitude, we find that static $BV$ magnitudes can be directly
estimated when both the mean quantities are measured. In
particular, either for F and FO pulsators, we derive that static
$BV$ values (see Table 2 for the other bands) can be approximated as

$$M_V=-0.345(M_V)+1.345<M_V>\eqno(15)$$
$$M_B-M_V=0.488[(M_B-M_V)]+0.497[<M_B>-<M_V>]\eqno(16)$$

\noindent with an intrinsic uncertainty of $\pm$0.01 mag.

\section{The evolutionary connection}

At the beginning of this paper we mentioned that the ``theoretical
route'' to the calibration of the $M_V$(RR) versus [Fe/H] relation,
as inferred by HB evolutionary models, suggests brighter
magnitudes than empirical estimates, at fixed [Fe/H].
Unfortunately, even the amount of this over-luminosity is still
matter of controversy.

For the sake of the following discussion, let us briefly summarize
the basic ingredients which are needed for a reliable ``theoretical
route'':

\begin{enumerate}

\item   Trustworthy HB models computed with the most recent updating of
the relevant physics.

\item   Synthetic HB simulations (SHB) in order to estimate the evolutionary
        effects and the role of different HB populations.

\item   Realistic bolometric corrections and color-temperature transformations.

\item   Appropriate scaling between the metallicity $Z$ and the
        iron-to-hydrogen content [Fe/H].

\end{enumerate}

Such a ``shopping list'' should help the common reader to understand
the main reasons for the rather different theoretical predictions
on the $M_V$(RR)-[Fe/H] relation, as presented in the recent
literature (see Caputo et al. 2000 and references therein).

It is known that, due to somehow different input physics, current
updated HB models still provide slightly different luminosity
values (see, e.g., Castellani 2003). Moreover, different
bolometric corrections, as well as different assumptions
concerning the value of the overabundance of $\alpha$-elements
with respect to iron, are used by the various researchers. We add
that sometimes, in order to account for the luminosity of actual
(evolved) pulsators, the zero age horizontal branch (ZAHB)
luminosity at the RR Lyrae instability strip is artificially
increased, by a fixed quantity or as a function of metallicity,
instead of making use of SHB results.

Within such a complex scenario, it is of primary importance to carefully
investigate into current uncertainties of HB models. This will be done in
the following section where the HB models at $Z$=0.001
([$\alpha$/Fe=0]) computed
by Cassisi et al.  (2003, in preparation, 
hereafter C03), are compared with VDB results.

\subsection{The zero age horizontal branch}

According to a very common procedure, ZAHB stars are investigated  
starting from stellar structures with the same mass
of the He core and the same chemical composition of the stellar
envelope as in the RG progenitor at the He ignition, but with an
increased abundance of Carbon in the He core by about 5\% by mass
as a result of the flash.  The mass of the H-rich envelope and,
thus, the total mass of the model are regarded as a free parameter
governed by the unknown amount of mass loss. A reliable ZAHB model
is  eventually obtained by evolving such an initial structure for
about 1 Myr to allow equilibrium of CNO elements in the H burning
shell (see, e.g., Caputo, Castellani \& Wood 1978, Caputo, Castellani \& 
Tornambe' 1978, Castellani, 
Chieffi \& Pulone 1991).

The C03 horizontal branch models used in this paper implement
previous calculations presented by Cassisi et al. (1999) and
Caputo \& Cassisi (2002) and will not discussed here. Of
importance is that both C03 and VDB models adopt quite similar
values for the mixing length parameter ($l/H_p$=2.0 and 1.9,
respectively), no helium and heavy element diffusion, and
can be taken as
representative of the range in {\it absolute magnitude} spanned by
recent HB evolutionary calculations with updated input physics.

Figure 7 shows the mass and luminosity of ZAHB models near the RR
Lyrae instability strip plotted as a function of the effective
temperature. As for the VDB models, note that they have the same
$overall$ metallicity $Z$=0.001, although referring to different
chemical mixtures ([$\alpha$/Fe]=0, [Fe/H]=$-$1.31 and
[$\alpha$/Fe]=0.3, [Fe/H]=$-$1.54).

From the comparison of models at fixed effective temperature, one
finds evidence for three main points:

\begin{enumerate}

\item   C03 models are brighter by $\delta$log$L \sim$ 0.01 dex;

\item   C03 models are more massive by $\sim 0.02M_{\odot}$;

\item   increasing [$\alpha$/Fe] up to 0.3, which is the mean value
estimated for globular cluster stars with [Fe/H]$<-$0.6 (see
Carney 1996), leaves quite unvaried the value of mass and
luminosity. In other words, at least in the range of RR Lyrae
effective temperature and with no significantly large
[$\alpha$/Fe] values, either the mass and the luminosity of
pulsators are expected to depend on the overall metallicity $Z$,
quite independently of the internal ratio between $\alpha$ and
heavy elements (see also Salaris, Chieffi \& Straniero 1993).

\end{enumerate}

As a first step, in order to gain some insight into the predicted
distribution of ZAHB pulsators, we evaluate for each model the
predicted FOBE and FRE, by inserting the mass and luminosity of
the model into eq. (5) and eq. (6), respectively, but with a
modified zero-point accounting for the mixing length parameter
adopted in the quoted evolutionary computations. In this way,
synthetic ZAHB pulsators are easily identified with the models for
which holds FOBE $\ge$ log$T_e \ge$ FRE.

As shown in Fig. 8 for C03 computations, one derives that the mass
of predicted ZAHB pulsators with $Z$=0.001 should vary from $\sim$
0.65 (FOBE) to $\sim 0.67M_{\odot}$ (FRE), and accordingly the
luminosity increases from log$L/L_{\odot}$=1.67 to 1.69. As for
VDB ZAHB models, we get mass in the range of $\sim$ 0.63 to
0.65$M_{\odot}$ and luminosity from log$L/L_{\odot}$=1.66 to 1.68.
It is worth mentioning that, since mass and luminosity vary in the
same direction, both sets of computations turn out to agree each
other as the effective temperature and the period of ZAHB
pulsators are concerned. We estimate that the effective
temperature of ZAHB pulsators varies from log$ T_e\sim$ 3.85
(FOBE) to $\sim$ 3.79 (FRE), while the predicted periods at the
edges of the instability strip  are log$P^{FO}$(FOBE)$\sim -$0.50
and log$P^F$(FRE)$\sim -$0.15, that are in reasonable agreement
with the observed color and period distributions of RR Lyrae stars
in globular clusters with well populated HB (see Paper I).

\subsection{Post ZAHB evolution and synthetic HB}

Since the first pioneering work by Rood (1973), the major way of
interpreting the properties of observed HB and RR Lyrae stars
deals with synthetic horizontal branches (SHB) constructed on the
basis of HB evolutionary tracks. The literature of the last three
decades is rather abundant (readers are referred to Ferraro et al.
(1999), Demarque et al. (2000) and references therein). However,
until now we have no knowledge of synthetic simulations as based
on either recent up-to-date HB models $and$ pulsation theory (i.e.
nonlinear, convective models). For this reason, an atlas of SHBs
with various metal content has been planned (Paper IV) with the
purpose to determine a series of parameters (HB morphology, RR
Lyrae luminosity, period distribution, etc) which are needed to
realistic discussion of observed HB and RR Lyrae stars in globular
clusters.

Figures 9 and 10 illustrate the results of two of these SHBs at
$Z$=0.001: specifically, those adopting a total number of HB stars
N(HB)=400, a mean mass of HB stars $M$(HB)=0.65 and
0.69$M_{\odot}$, and a mass dispersion factor by 0.02$M_{\odot}$
(see Demarque et al. 2000). Following the procedure described for
ZAHB models, synthetic HB stars hotter (cooler) than their
predicted FOBE (FRE) are taken as blue (red) stable stars, while
those with FOBE$\ge$ log$T_e\ge$ FRE  are identified as pulsating
structures. In this way, the predicted numbers of blue (B),
variable (V) and red (R) stars are derived, leading to the
(B-R)/(B+V+R) ratio (Lee 1990) labelled in the two panels. In the
figures, the solid  line is the ZAHB, while the dashed vertical
ones depict the predicted edges of the RR Lyrae instability strip.

Besides the well known correlation between the average mass
$M$(HB) and the HB morphology (lower is the mean mass of HB stars
bluer their color distribution, everything else being constant),
we should mark three points:
\begin{enumerate}

\item   synthetic pulsators are brighter and, on average, slightly
        more massive than ZAHB ones. Note that the brightest stars 
	populating the RR Lyrae instability strip are also the less 
	massive ones;

\item   the lower envelope of the evolved star luminosities is slightly
        brighter than the ZAHB luminosity level (see also Ferraro 
	et al. 1999). Note that this may depend on the above definition of 
	``ZAHB'' (Sect. 3.1.);

\item   the average mass of the variable stars is quite constant,
        $<M(RR)>$=0.67$\pm$0.02$M_{\odot}$, in the quoted range of $M$(HB)
        and (B-R)/(B+V+R) ratio. We show in Table 4 that,
    as long as the fraction of variables is statistically significant
        [V/(B+V+R)$\ge$0.20], the above value for the pulsator average
        mass holds for (B-R)/(B+V+R) ratio in the range of $\sim$+0.55 to
        $\sim-$0.45, whereas for bluer or redder HB distributions the SHB
        results suggest $<M(RR)>$=0.66$\pm$0.02$M_{\odot}$ and
        0.68$\pm$0.02$M_{\odot}$, respectively. With small numbers of
        variables, i.e. very blue or very red HB distributions, no statistically
        reliable information on the average mass $<M(RR)>$ can be derived.

\end{enumerate}

Let us now investigate the distribution of the synthetic HB
pulsators, namely of the evolved stellar structures satisfying the
condition FOBE $\ge$ log$T_e \ge$ FRE, in the $M_V$-log$P$ plane.
We present the case at $M$(HB)=0.67$M_{\odot}$ (HB index $\sim$
0.10, as observed in M3), but quite similar results  are obtained
with $M$(HB)=0.65 and 0.69$M_{\odot}$. As already mentioned, we
adopt CGKa,b bolometric corrections.

The left panel in Fig. 11 deals with the short-period side of
synthetic first overtone pulsators [periods from eq. (1b)]. The
solid line is the resulting FOBE  of the synthetic distribution
(see the labelled relation), while the dashed lines depict its
intrinsic dispersion [see eq. (5)] of $\pm$ 0.034 mag. In the
right panel, which refers to the long-period side of fundamental
periods [periods from eq. (1a)], the solid and dashed lines depict
the predicted FRE (see labelled relation) and its intrinsic
uncertainty by $\pm$ 0.04 mag [see eq. (6)], respectively. The
horizontal line drawn in the two panels depicts the lower envelope
magnitude $M_{V,le}$=0.55 mag of the synthetic pulsator
distribution.

A question that should be soon asked is ``how far the above
predictions on the FOBE and FRE of evolved pulsators are
model-dependent?''. Bearing in mind the main differences between
C03 and VDB models (see Fig. 7), we have generated ``VDB-like''
synthetic populations by simply decreasing mass and luminosity of
our HB models by 0.02$M_{\odot}$ and 0.01 dex, respectively, and
maintaining CGKa,b bolometric corrections. As a result, we get
that either FOBE and FRE magnitudes should be increased by $\sim$
0.03 mag, at fixed period, as well as that the absolute magnitude
of the lower envelope is  $M_{V,le}$=0.58 mag.

\section{The RR Lyrae population in M3}

We are now in the position to undertake an exhaustive comparison
between the pulsational predictions presented in the previous
sections and the observed properties of RR Lyrae stars in M3.
Taking advantage of the extensive set of $BV$ photometric data
provided by Corwin \& Carney (2001), we select\footnote{Variables
with Blashko effect, blend or less precise periods are excluded.}
a fiducial sample of variables with quite well-measured periods,
mean magnitudes and amplitudes. In the following discussion, we
will adopt $E(B-V)$=0.

Figure 12 shows the variables (different pulsation modes follow
Corwin \& Carney (2001) classification) plotted in the
color-magnitude diagram, according to magnitude-averaged (upper
panel) and intensity-averaged (lower panel) quantities. As a
matter of reference, we draw a line at $V$=15.8 mag. At first
sight, the $ab$-type magnitude plotted in the upper panel are on
average fainter  than in the lower panel. Moreover, the
magnitude-averaged color of the bluest $ab$-type variables, is
evidently redder than the intensity-averaged one. As an example,
V77 shows $(B-V)$= 0.279 mag and $<B>-<V>$=0.223 mag.

And indeed, we show in Fig. 13 that, with the only exception of
variable V82 (filled triangle), the behavior of the measured
difference between the empirical mean quantities follows the
predictions presented in Fig. 5 and Fig. 6 faithfully. On this
ground, eq. (15) and eq. (16) are used to get static magnitudes
and colors, with an error of $\pm$ 0.01 mag, giving us the
opportunity to compare the predictions presented in the previous
sections with observed data.

Let us firstly consider the predicted relations which are
independent, or slightly dependent, of mass and luminosity.

{\it a) The Color-Amplitude relation}

Figure 14 shows the visual amplitude of $ab$-type variables in M3
as a function of the static $B-V$ color, in comparison with the
predicted $CA_{BV}$ relations at $l/H_p$=1.5 (solid line) and 2.0
(dotted line). Bearing in mind the intrinsic dispersion ($\pm$
0.02 mag in color, at fixed amplitude) of the predicted relations,
and allowing for photometric errors, the agreement between theory
and observations is reasonably good, disclosing also a quite
interesting feature: at the larger amplitudes, the observed data
are not allowing a firm choice between the two adopted values of
the mixing-length parameter, but moving towards redder colors the
measured amplitudes are clearly pointing to $l/H_p$=2.0.

{\it b) The Color-Period diagram}

Figure 15 shows the RR Lyrae stars in M3 (symbols as in Fig. 12)
plotted in the $B-V$ versus period diagram, in comparison with the
predicted FOBE [eq. (9)] and FRE [eq. (10)] at $l/H_p$=1.5 (solid
line) and 2.0 (dotted line).  As for the mass of the variables, we
adopt the value $<M(RR)>$=0.67$\pm 0.02M_{\odot}$ suggested by the
SHB results presented in Fig. 9, bearing in mind that the color of
the predicted edges depends very little on the mass. Taking into
account the intrinsic uncertainty ($\pm$ 0.02 mag) of the
predicted color at fixed period, the  data plotted in Fig. 15 show
that the reddest $ab$-type variables closely match
the predicted FRE with $l/H_p$=2.0, excluding the case
$l/H_p$=1.5, unless a {\it negative} reddening is adopted.

Concerning  
the bluest $c$-type variables, excluding V177 and V178, they are
in a marginal agreement with the theoretical FOBE with
$l/H_p$=1.5, whereas the case 
$l/H_p$=2.0 appears significantly too red. 
As for the location of V177 and V178, one has three possible
choices, provided that the measured color is correct: 1) the variables are
not {\it bona fide} FO pulsators (second overtones ?); 2) the
variables are dramatically more massive ($\sim$ 50\%) than the
bulk of M3 variables [see eq. (9)]; 3) the actual FOBE is bluer than predicted
with $l/H_p$=1.5. As for the last case, we estimate that adopting
a mixing-length parameter $l/H_p \sim$ 1.1 will move blueward the
FOBE (see Caputo et al. 2000) by the needed shift $\sim$ 0.02 mag,
at fixed period.

Let us now proceed with predicted relations involving mass and
luminosity of the variables, beginning with those
which are independent of the edges of the instability
strip, and then of the adopted mixing-length parameter.

{\it c) Period-Magnitude-Color relation}

According to eq. (2a) and eq. (2b), one may constrain the mass of
RR Lyrae stars for each given assumption on the distance modulus
$DM$ or, alternatively, one can derive the distance modulus for
each given value of the mass. 

Adopting two
alternative values for the cluster distance modulus, $DM$=14.9 and
15.2 mag, taken as roughly representative of the range of current
estimates (see Saad \& Lee 2001), we
show in Fig. 16 the reduced fundamentalized 
period (log$P^F$+0.34$M_V$) of RR Lyrae stars in M3 (symbols as in
Fig. 12) as a function of the static $B-V$ color. 
As a whole, the data reasonably
follow the predicted slope at constant mass (solid line), even though  
the observations would suggest a slope of 1.05 which is less than   
the predicted one.  
It seems of interest to note that previous simulations (Rood 1990)
do show that observational color errors do make the observed slope less than 
the theoretical one. 

According to the predicted relations, increasing $DM$ from 14.9 to 15.2 mag, 
the average
mass of RR Lyrae stars increases from $\sim$ 0.54 to
0.77$M_{\odot}$, largely encompassing the value $<M(RR)>$=0.67$\pm
0.02M_{\odot}$ suggested by the SHB results presented in Fig. 9.
As a fact, if the {\it evolutionary}  mass is adopted,
then the resulting distance
modulus is $DM(PMC)$=15.09$\pm$0.08 mag.

As a fact, if the {\it evolutionary}  mass is adopted, 
then the resulting distance
modulus is $DM(PMC)$=15.09$\pm$0.08 mag.

Interesting enough, inspection of  Fig. 16 shows that the reduced
period of V177 and V178 is longer with respect to the general
behavior of M3 $c$-type variables. According to eq. (2b), this
difference should suggest that the mass of these variables is
smaller ($\sim$ 20\%) than the remaining $c$-type stars, at
variance with the mass variation we needed in Fig. 15 to bring
their color in agreement with the predicted FOBE. On the other hand, if a
decrease of the mixing-length parameter can be invoked in Fig. 15
to move blueward the predicted FOBE, this is without significance
for what the $PMC$ relation is concerned. Furthermore, we could
also exclude that these variables are second-overtone pulsators
since in such a case one expects shorter periods than
first-overtone pulsation, at fixed color (i.e. $T_e$). In
conclusion, the pulsation models would suggest that the measured
color of V177 and V178 should be increased by $\sim$ 0.03 mag, as required by
 both the FOBE location and the $PMC$ relation.

{\it d) Near-infrared period-magnitude relation}

Proceeding in our analysis, we compare measured $K$ magnitudes of
RR Lyrae stars in M3 with the predicted near-infrared
period-magnitude relation. Using Longmore at al. (1990)
near-infrared photometry and fundamentalized periods, we show in
Fig. 17 that the data arrange quite well along  the predicted
slope at constant mass and luminosity (solid line). Since the
predicted $M_K$ magnitude is slightly dependent on the intrinsic
luminosity (see lower panel in Fig. 2), we adopt
log$L$=1.75$\pm$0.15 as a quite safe luminosity range for RR Lyrae
with $Z$=0.001. Then, taking again
$<M(RR)>$=0.67$\pm$0.02$M_{\odot}$, we derive
$DM(PM_K)$=15.08$\pm$0.06 which is very close to the $PMC$ result.

{\it e) Period-magnitude-amplitude relation}

As a further independent method to estimate the distance modulus,
let us consider the period-magnitude-amplitude ($PMA$) of
fundamental pulsators. Note that now one has different results,
depending on the mixing-length  parameter [see eq. (13) and eq.
(14)], for each given mass. Adopting again $<M(RR)>$=0.67$\pm
0.02M_{\odot}$, we shown in Fig. 18 that fitting observed data to  
the predicted $PMA$ relations with $l/H_p$=1.5 (left panel) 
and $l/H_p$=2.0 (right panel) yields 
$DM(PMA)$=14.98$\pm$0.08 mag and 15.09$\pm$0.07 mag, 
respectively. 

{\it f) Edges of the instability strip}

Figure 19 finally illustrates the comparison between the observed
distribution of RR Lyrae stars in the $V$-log$P$ plane and the
predicted edges of the instability strip, as inferred by SHB
simulations with $l/H_p$=2.0 (see Fig. 11). 
One has that a nice
agreement between predictions and observations is attained
adopting $DM$=15.04 mag, even though the period of a couple of
$c$-type variables turns out be slightly shorter than the predicted 
limit. 
On the other hand, if the distance modulus is
slightly decreased ($DM$=15.0 mag) in order to leave no $c$-type  
variables in the hot stable region, then a few $ab$-type variables
would have slightly longer period than the predicted FRE.

As far a decrease of the adopted mixing-length parameter is
concerned, one has to consider that at $l/H_p$=1.5 the predicted
edges becomes brighter by 0.05 mag (FOBE) and fainter than 0.14
mag (FRE), at constant period. As a consequence, a distance
modulus $DM$=15.05 mag is the right value to fit the observed
distribution of $c$-type variables, whereas $ab$-type variables
would require $DM$=14.90 mag. 

We summarize in Table 4 the distance moduli inferred by the
different pulsational methods discussed so far. One can conclude
that the assumption $l/H_p$=1.5 yields rather discordant results,
while a quite close agreement among the various approaches is
reached with $l/H_p$=2.0. However, we have shown in Fig. 15 that adopting
such a large value of the mixing-length parameter the predicted
FRE fully accounts for the observed colors of $ab$-type variables,
whereas the predicted FOBE is too red in comparison with the
observed colors of $c$-type variables. On this issue, it seems
interesting to mention a recent paper by Bono, Castellani \&
Marconi (2002) dealing with the comparison of observed light
curves of bump Cepheids in the Large Magellanic Cloud with
theoretical predictions. In that paper it is shown that for
short-period variables a nice agreement is attained with
$l/H_p$=1.5, whereas long-period Cepheids, which are located close
to the red edge of the pulsation region, require a higher value
($l/H_p$=1.8), suggesting that the mixing-length parameter
increases when the effective temperature decreases, and the
convective layer becomes thicker.

Eventually, we adopt that the mixing-length parameter increases
from FOBE ($l/H_p$=1.5) to FRE ($l/H_p$=2.0). As a result, we
derive that the {\it pulsational} distance modulus of M3, as given 
by the weighted
mean over the data listed in Table 4, is $DM$=15.07$\pm$0.05 mag,
at least adopting RR Lyrae masses from C03 HB models and CGKa,b
bolometric corrections.

\section{Discussion and final remarks}

In the previous section we have presented evidence that
theoretical relations based on pulsating models with $Z=0.001$,
collated with information on the average mass of RR Lyrae stars
inferred by SHB computations, provide a $pulsational$ route to
accurate distance determinations for globular clusters with well
populated horizontal branch.
In particular, adopting $<M(RR)>$=0.67$\pm$0.02$M_{\odot}$ from
SHB calculations based on
C03 HB models, we derived
predictions for fundamental (F) or first-overtone (FO)
RR Lyrae stars in globular clusters with (B-R)/(B+V+R) ratio from
$\sim$+0.55 to $\sim-$0.45.

The comparison with RR Lyrae stars in M3 
($BV$ data by Corwin \& Carney 2001, $K$ data by Longmore et al.
1990) 
discloses that the
pulsating models support the large value of the mixing-length
parameter ($l/H_p$=1.9-2.0) adopted in updated evolutionary
computations to fit observed Red Giant Branches, even though they
suggest that the value of such a parameter may increase from the
blue to the red edge of the instability region.
On this basis, adopting CGKa,b bolometric corrections and 
color-transformations, 
we eventually get a {\it pulsational} distance modulus 
$DM$(M3)=15.07$\pm$0.05 mag, as given by
the weighted mean over the results provided by the different
approaches (see Table 4).

In order to compare this result with the classical 
{\it evolutionary route},  
let us go back to Fig. 10. There, we  
have shown that the lower envelope of the absolute magnitudes of 
synthetic
pulsators based on C03 HB models is at $M_{V,le}$=0.55 mag. On the other 
hand,  we show in
Fig. 20 that the actual lower envelope for the M3 variables 
is $V$=15.7 mag (solid line), thus providing an {\it evolutionary} 
distance modulus $DM$=15.15 mag which is  0.08$\pm$0.05 mag longer 
than the {\it pulsational} value. 

Before discussing such a small but equally disturbing discrepancy
between the pulsational  and evolutionary approach, 
let us mark that
it {\it does not depend on the adopted HB models, nor on the
adopted bolometric correction}. In fact, were the VDB models used,
but maintaining CGKa,b bolometric corrections, then the average
mass of RR Lyrae pulsators would be 0.65$\pm$0.02$M_{\odot}$ (point 2 in
Sect. 3.1), and consequently the distance modulus based on $PMC$,
$PMA$ and $PM_K$ relations should be decreased by $\sim$ 0.02 mag, on 
average.
On the other hand, the predicted FOBE and FRE magnitudes, at fixed
period, increases by 0.03 mag (see Fig. 11). As a consequence, the
pulsational distance modulus of M3 becomes $DM$=15.05$\pm$0.05
mag. Since the lower envelope of the VDB-like pulsators is at
$M_{V,le}$=0.58 mag (see Fig. 11), one derives again that the 
pulsational approach yields a distance modulus larger by 
0.07$\pm$0.05 mag with respect to the 
evolutionary predictions. 

As for the effects of different bolometric corrections, it is worth mentioning 
that there is a significant difference between CGKa,b values and
those adopted in VDB models, in the sense that the former give
absolute magnitudes brighter by $\sim$ 0.05-0.06 mag, at fixed luminosity
and effective temperature. However,
such a difference obviously modifies the absolute magnitudes, and
then the derived distance modulus, but leaves unvaried relative
values, mainly the above discrepancy between pulsational and evolutionary 
distance moduli. 

In summary, the distance modulus of M3, as based on pulsational
models and mass constrained by stellar evolution theory, ranges
from 15.00 mag (VDB computations and bolometric corrections) to
15.05 mag (VDB computations and CGKa,b bolometric corrections) to
15.07 mag (C03 computations and CGKa,b bolometric corrections),
all with a total error of $\pm$ 0.05 mag. It follows that recent
HB computations with updated physics do yield consistent results,
{\it provided that similar bolometric corrections are used}.
Conversely, current  uncertainty on the $BC$ scale appears still
too large, requiring some firm solutions.

Independently of the adopted $BC$ correction, the pulsational
approach would suggest that current updated HB models are
over-luminous by 0.08$\pm$0.05 mag. However, a clear way to reduce
the HB luminosity is offered by the diffusion of helium and
metals. As shown by Cassisi et al. (1999), if element diffusion is
properly taken into account, then the luminosity of HB models at
the RR Lyrae gap decreases by 0.03-0.04 mag, reducing  the
discrepancy between pulsational and evolutionary distances to
$\sim$ 1$\sigma$.

\section{Acknowledgments}
We thank our referee, B. Rood, for very useful comments and suggestions.
Financial support for this work was provided by MIUR-Cofin 2000,
under the scientific project ``Stellar Observables of Cosmological
Relevance''. Model computations made use of resources granted by
the ``Consorzio di Ricerca del Gran Sasso'', according to Project 6
``Calcolo Evoluto e sue Applicazioni (RSV6) - Cluster C11/B''.

\pagebreak

\clearpage

\clearpage
\begin{table}[widht=1\twidht]
\begin{center}
\caption{Selected results for fundamental pulsating models at
$Z$=0.001, $Y$=0.24, $M=0.65M_{\odot}$, log$L/L_{\odot}$=1.61 and
$l/H_p$=1.5.}
\vspace{0.3 cm}
\begin{tabular}{lcccccc}
\hline
T$_{e}$  && $M_U$ & $M_B$ & $M_V$ & $M_I$ & $M_K$\\
\hline
 6000(FRE) && 1.242 & 1.265 & 0.814    &  0.179   & -0.586 \\
 6100 && 1.207 & 1.229 & 0.802    &  0.194    & -0.534 \\
 6300 && 1.149 & 1.164 & 0.781    &  0.225     & -0.431 \\
 6500 && 1.104 & 1.104 & 0.761    &  0.258     & -0.332 \\
 6700 && 1.067 & 1.050 & 0.744    &  0.292     & -0.233 \\
 6900(FBE) && 1.037 & 1.001 & 0.729    &  0.327     & -0.139 \\
\hline
\hline
\hline
 T$_e$ & P &  $<M_U>$ & $<M_B>$ & $<M_V>$ & $<M_I>$ &  $<M_K>$ \\
\hline
6000 & 0.6598  &  1.246  & 1.265  & 0.819 &   0.188 &  -0.575 \\
6100 & 0.6254  &  1.225  & 1.241  & 0.812 &   0.200 &  -0.540 \\
6300 & 0.5626  &  1.163  & 1.168  & 0.787 &   0.239 &  -0.436 \\
6500 & 0.5061  &  1.119  & 1.104  & 0.768 &   0.268 &  -0.342 \\
6700 & 0.4567  &  1.077  & 1.043  & 0.759 &   0.317 &  -0.239 \\
6900 & 0.4137  &  1.037  & 1.010  & 0.770 &   0.376 &  -0.137 \\
\hline
\hline
\hline
T$_e$ & P & $(M_U)$ & $(M_B)$ & $(M_V)$ & $(M_I)$ &  $(M_K)$ \\
\hline
6000 &  0.6598 & 1.253 &  1.273 & 0.823 &   0.190 &  -0.573 \\
6100 &  0.6254 & 1.238 &  1.257 & 0.820 &   0.204  & -0.537 \\
6300 &  0.5626 & 1.187 &  1.198 & 0.804 &   0.239  & -0.433 \\
6500 &  0.5061 & 1.168 &  1.167 & 0.805 &   0.283   &-0.338 \\
6700 &  0.4567 & 1.164 &  1.156 & 0.824 &   0.341  & -0.234 \\
6900 &  0.4137 & 1.184 &  1.180 & 0.869 &   0.412  & -0.131 \\
\hline
\hline
\hline
T$_e$ & P &    $A_U$ & $A_B$ & $A_V$ & &  $A_K$ \\
\hline
6000 &  0.6598 &     0.438 &   0.466 &  0.349 & &    0.182 \\
6100 &  0.6254 &     0.636 &   0.690 &  0.528 & &   0.269 \\
6300 &  0.5626 &     0.850 &   0.958 &  0.740 & &   0.285 \\
6500 &  0.5061 &     1.064 &   1.206 &  0.963 & &   0.331 \\
6700 &  0.4567 &     1.552 &   1.703 &  1.335 & &   0.334 \\
6900 &  0.4137 &     1.967 &   2.003 &  1.557 & &   0.408 \\
\hline
\end{tabular}
\end{center}
\end{table}
\clearpage
\begin{table}[widht=0.8twidht]
\begin{center}
\vspace{0.3 cm}
\caption{As in Table1, but for first-overtone pulsating models}
\vspace{0.3 cm}
\begin{tabular}{lcccccc}
\hline
T$_{e}$ & & $M_U$ & $M_B$ & $M_V$ & $M_I$ &  $M_K$\\
\hline
6600(FORE) & &    1.085 &    1.076 &    0.752 & 0.275 &    -0.282 \\
6700 & &    1.068 &    1.050 &    0.744 & 0.293 &    -0.233 \\
6800 & &    1.051 &    1.024 &    0.736 & 0.310 &   -0.185 \\
6900 & &    1.037 &    1.000 &    0.728 & 0.327 &   -0.139 \\
7000 & &    1.024 &    0.977 &    0.722 & 0.346 &   -0.092 \\
7100 & &    1.012 &    0.955 &    0.716 & 0.364 &    -0.046 \\
7200(FOBE) & &    1.001 &    0.934 &    0.711 & 0.383 &     0.000 \\
\hline
\hline
\hline
T$_e$ & P &  $<M_U>$ & $<M_B>$ & $<M_V>$ & $<M_I>$  & $<M_K>$ \\
\hline
6600 &   0.3558 & 1.100 & 1.091 & 0.762 &   0.277 & -0.296 \\
6700 &   0.3362 & 1.083 & 1.063 & 0.754 &   0.295 &  -0.251 \\
6800 &   0.3211 & 1.066 & 1.036 & 0.747 &   0.315 & -0.203 \\
6900 &   0.3055 & 1.052 & 1.011 & 0.742 &   0.336 & -0.155 \\
7000 &   0.2921 & 1.036 & 0.986 & 0.738 &   0.357 &  -0.109 \\
7100 &   0.2779 & 1.020 & 0.962 & 0.734 &   0.378 &  -0.059 \\
7200 &   0.2663 & 1.008 & 0.941 & 0.727 &   0.396 &  -0.007 \\
\hline
\hline
\hline
T$_e$ & P & $(M_U)$ & $(M_B)$ & $(M_V)$ & $(M_I)$ &  $(M_K)$ \\
\hline
6600 & 0.3558 & 1.115 & 1.111 & 0.775 &  0.282 &   -0.295 \\
6700 & 0.3362 & 1.109 & 1.097 & 0.775 &  0.304  &  -0.249 \\
6800 & 0.3211 & 1.101 & 1.082 & 0.776 &  0.326 &   -0.201 \\
6900 & 0.3055 & 1.092 & 1.066 & 0.776 &  0.349  &  -0.153 \\
7000 & 0.2921 & 1.085 & 1.052 & 0.778 &  0.372 &   -0.106 \\
7100 & 0.2779 & 1.073 & 1.033 & 0.776 &  0.394  & -0.057 \\
7200 & 0.2663 & 1.045 & 0.992 & 0.757 &  0.407 &   -0.005 \\
\hline
\hline
\hline
T$_e$ & P &   $A_U$ & $A_B$ & $A_V$ & &  $A_K$ \\
\hline
6600 &   0.3558 &    0.521 & 0.599 &    0.473 & &  0.189 \\
6700 &   0.3362 &    0.680 & 0.778 &    0.618 & & 0.217 \\
6800 &   0.3211 &    0.830 & 0.962 &    0.756 & & 0.218 \\
6900 &   0.3055 &    0.891 & 1.035 &    0.807 & & 0.216 \\
7000 &   0.2921 &   1.039 & 1.192 &    0.919 & &  0.217 \\
7100 &   0.2779 &    1.045 & 1.187 &    0.916 & & 0.191 \\
7200 &   0.2663 &    0.843 & 0.971 &    0.745 & & 0.149 \\
\hline
\end{tabular}
\end{center}
\end{table}

\begin{table}
\begin{center}
\vspace{0.3 cm}
\caption{Average mass $<M(RR)>$ of synthetic RR Lyrae pulsators 
varying the
mean mass $<M(HB>$ of HB stars, as inferred by SHB computations
based on C03 models with $Z$=0.001 and $\sigma_{M(HB)}$=0.02
$M_{\odot}$.
The
HB-type and the fraction of variable stars $f(RR)$
are also listed. The predicted
mass spread of statistically significant numbers
of RR Lyrae pulsators ($f(RR)\ge$ 0.20)
is of the order of $\sim$ 0.02$M_{\odot}$. For very blue or
very red HB types ($<M(RR)>$ values in italics), no reliable
average mass can be derived.}
\begin{tabular}{cccc}
\hline
$<M(HB>$ & HB-type & $f(RR)$ & $<M(RR)>$\\
($M_{\odot}$) &  &  & ($M_{\odot}$)\\
\hline
0.62 & +0.93 & 0.04 & {\it 0.64}\\
0.64 & +0.72 & 0.18 & {\it 0.66}\\
0.65 & +0.55 & 0.27 & 0.667\\
0.66 & +0.38 & 0.35 & 0.669\\
0.67 & +0.10 & 0.37 & 0.670\\
0.68 & $-$0.19 & 0.40 & 0.673\\
0.69 & $-$0.45 & 0.30 & 0.675\\
0.70 & $-$0.72 & 0.21 & 0.679\\
0.72 & $-0.95$ & 0.04 & {\it 0.69}\\
\hline
\end{tabular}
\end{center}
\end{table}

\begin{table}
\begin{center}
\vspace{0.3 cm}
\caption{Distance modulus (mag) of M3 as inferred from the different
 pulsational methods described in the text}
\begin{tabular}{lcc}
\hline
Method & $DM$($l/H_p$=1.5) &  $DM$($l/H_p$=2.0)\\
\hline
$PMC$  &  15.09$\pm$0.08 & 15.09$\pm$0.08\\
$PMA$  &  14.98$\pm$0.10 & 15.09$\pm$0.08\\
$PM_K$ & 15.08$\pm$0.06 & 15.08$\pm$0.06\\
FOBE &  15.04$\pm$0.04 & 15.00$\pm$0.04\\
FRE  &  14.90$\pm$0.04 & 15.04$\pm$0.04\\
\hline
\end{tabular}
\end{center}
\end{table}
\clearpage



\figcaption{Period-magnitude-color relations of fundamental (F)
and first-overtone (FO) pulsators with the labelled mass and
luminosity. The plotted points are the models
which are separated 
by 100K in the effective temperature.}




\figcaption{Period-magnitude relations of fundamental (F) and
first-overtone (FO) pulsators. Symbols as in Fig. 1.}


\figcaption{Period-amplitude distribution of fundamental pulsators
at $l/H_p$=1.5 (left panel) and various mass and luminosity (symbols as
in Fig. 1). The right panel shows the predicted period-amplitude relations at
$l/H_p$=1.5 and 2.0 (see text).}


\figcaption{Color-amplitude distribution of of fundamental (F) 
pulsators at $l/H_p$=1.5 (left panel) and various mass and luminosity 
(ymbols as in Fig. 1). The right panel shows the predicted 
color-amplitude relations at
$l/H_p$=1.5 and 2.0 (see text).} 

\figcaption{Comparison between static and mean magnitudes of
fundamental (F) and
first-overtone (FO) pulsators.}


\figcaption{Comparison between static and mean colors of fundamental (F) and
first-overtone (FO) pulsators.}


\figcaption{Mass and luminosity of selected ZAHB models by C03 and
VDB as a function of the effective temperature.}

\figcaption{Mass and luminosity of selected ZAHB models by C03 as
a function of the difference log$T_e-$FOBE and log$T_e-$FRE. The arrows depict
the predicted limits of the instability region.}

\figcaption{Mass of synthetic HB stars as a function of the
effective temperature, normalized to FOBE, and varying
the average mass $<M(HB)>$. The
solid line is the ZAHB, while the dashed lines depict the
edges of the pulsation region.}

\figcaption{As in Fig. 8, but for the luminosity.}

\figcaption{Absolute visual magnitude of C03 synthetic HB
pulsators as a function of period. Left and right panel refers to
first-overtone and fundamental pulsators, respectively. The solid
lines (see the labelled relations) are the predicted edges of the
instability strip, while the dashed lines depict their intrinsic
dispersion.}

\figcaption{Color-magnitude diagram of RR Lyrae stars in M3. The
two panel refer to magnitude-weighted and intensity weighted
values. Different symbols refer to different pulsation modes.
The dashed line is drawn at 15.8 mag.}

\figcaption{Difference between observed mean magnitudes and mean colors for
RR Lyrae stars in M3. Symbols are the same as in previous figure apart from
the filled triangle which labels the only deviating variable V82.}

\figcaption{Color-amplitude diagram of
RR$_{ab}$ in M3 in comparison with 
the predicted relations under 
different assumptions on the mixing-length parameter.}

\figcaption{Color-period diagram of RR Lyrae stars in M3 in comparison
with the predicted edges of the instability strip at 
two 
different assumptions on the mixing-length parameter. 
Symbols as in Fig. 12. 
Variables V177 and V178 are
discussed in the text.}

\figcaption{Reduced period versus color for
RR Lyrae stars in M3 (symbols as in Fig. 12),
under two different assumptions on
the distance modulus. The period of $c$-type variables has 
been fundamentalized and the the solid line depicts
the predicted fundamental relation at constant mass.  
The resulting average mass  $<M(RR>$ of RR Lyrae stars for the labelled
distance modulus is given.}

\figcaption{Near-infrared magnitudes of RR Lyrae stars in M3 as a
function of period. The solid line is the predicted slope at constant mass 
and luminosity (see
text).}

\figcaption{Period-amplitude diagram of
RR$_{ab}$ in M3 in comparison with
the predicted relations (solid lines)
under different assumptions
on the mixing-length parameter.
The labelled distance moduli are
the resulting values adopting an average mass
$<M(RR)>$=0.67$\pm$0.02$M_{\odot}$}

\figcaption{RR Lyrae stars in M3 in the $V$-log$P$ plane in comparison 
with the predicted edges of the instability strip at 
$l/H_p$=2.0, as corrected with the labelled distance modulus.}

\figcaption{Color-magnitude diagram of RR Lyrae stars in M3.
The solid line is the observed magnitude of the
lower envelope.}

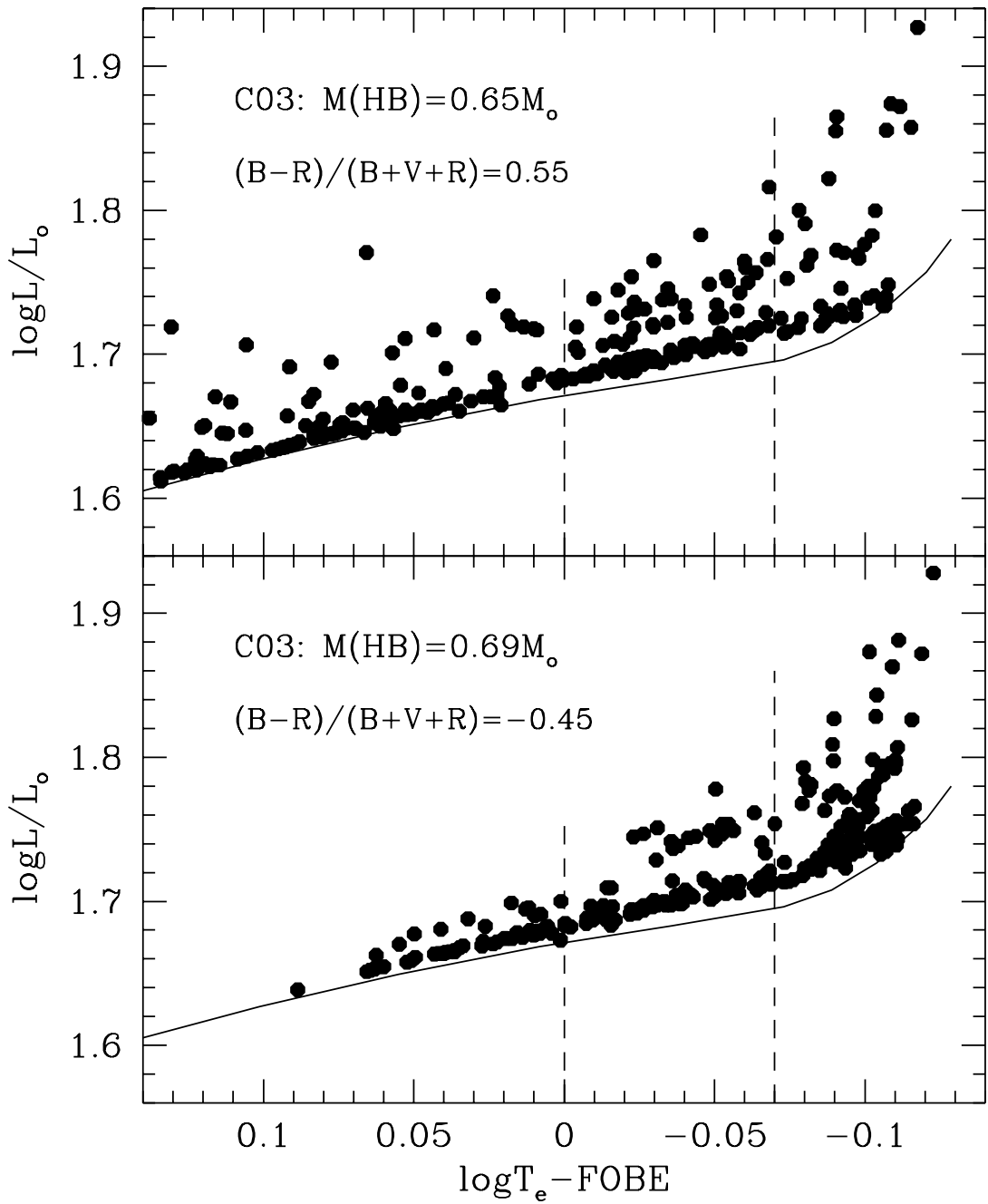
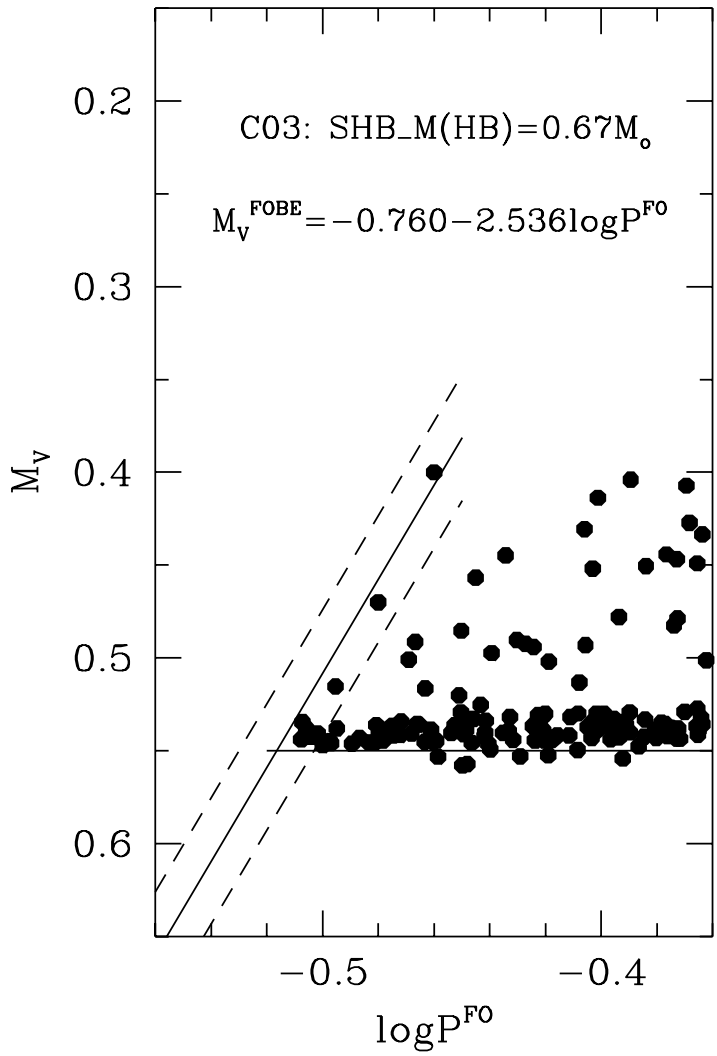
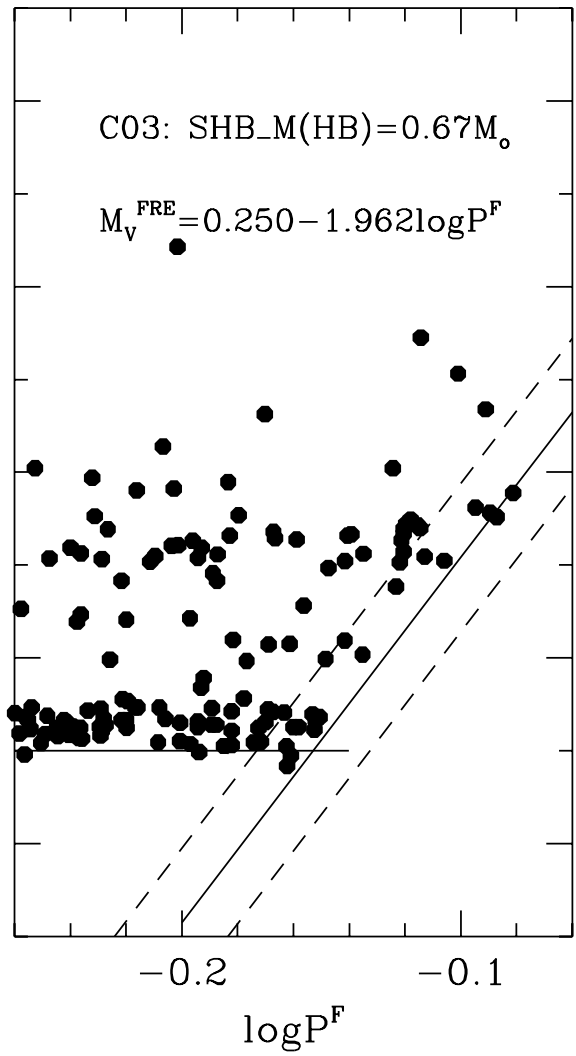
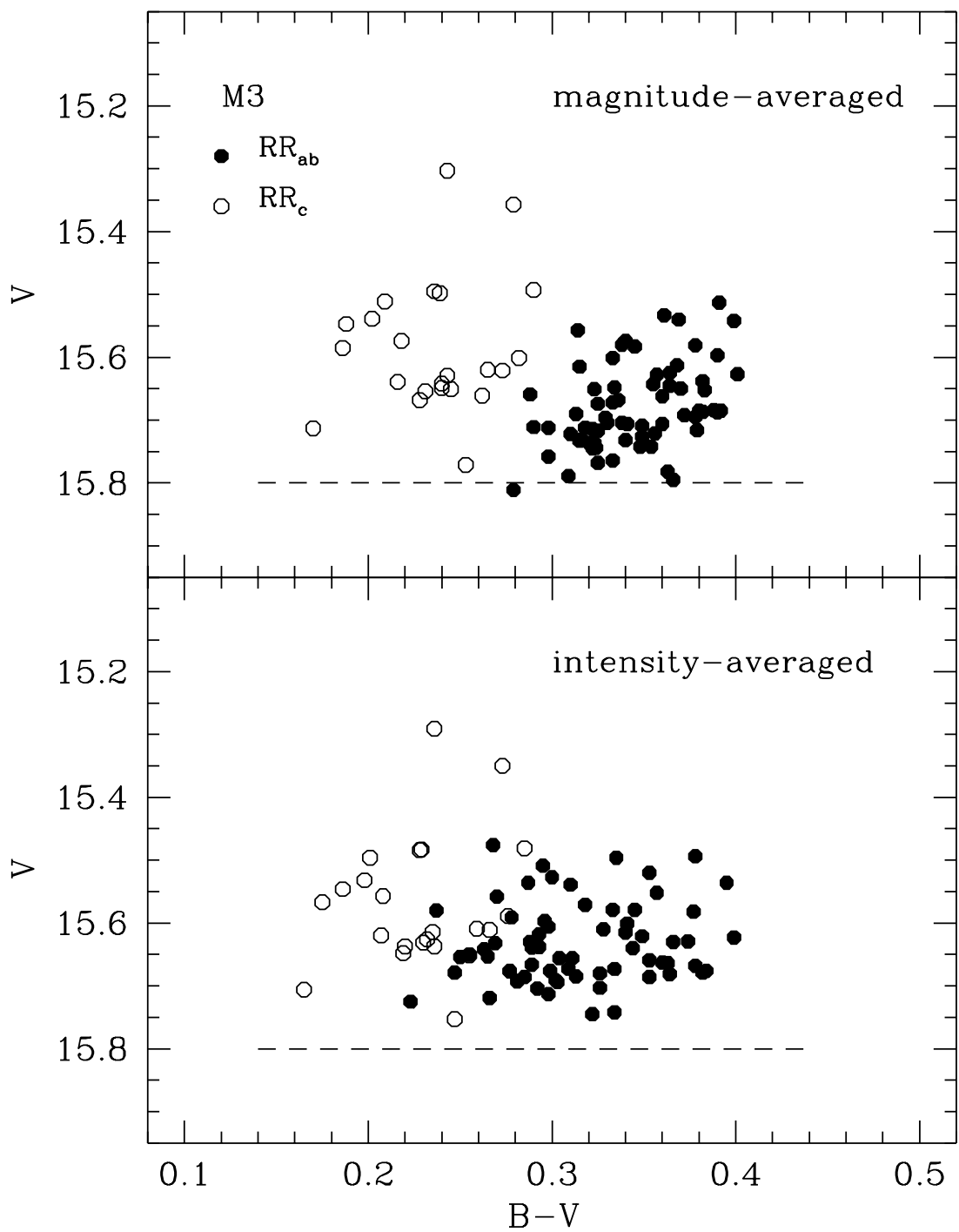
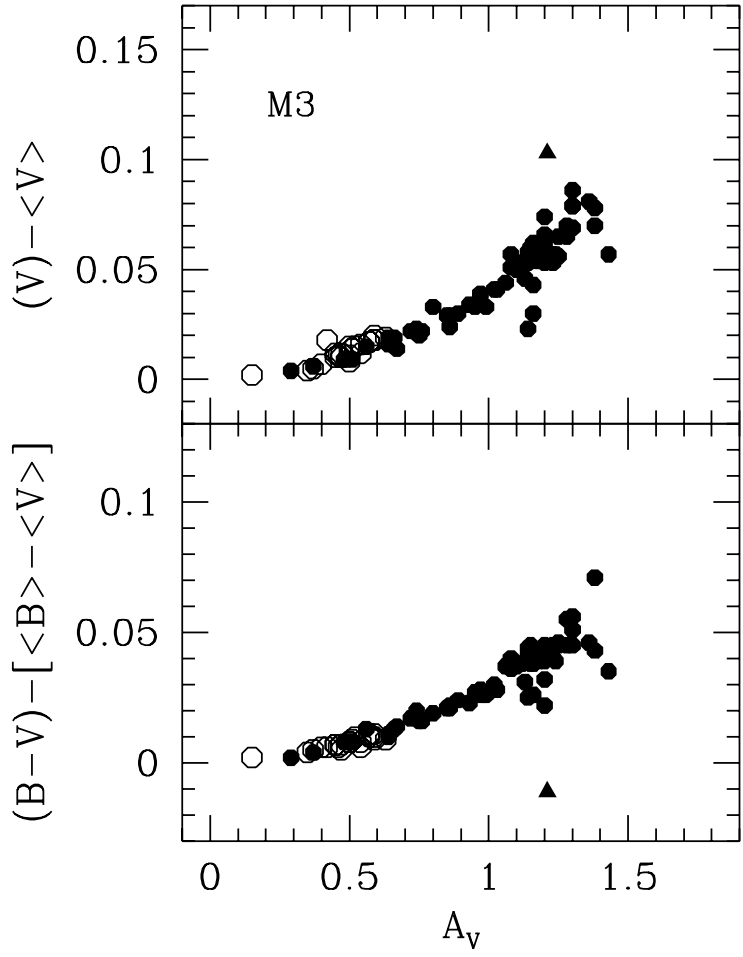
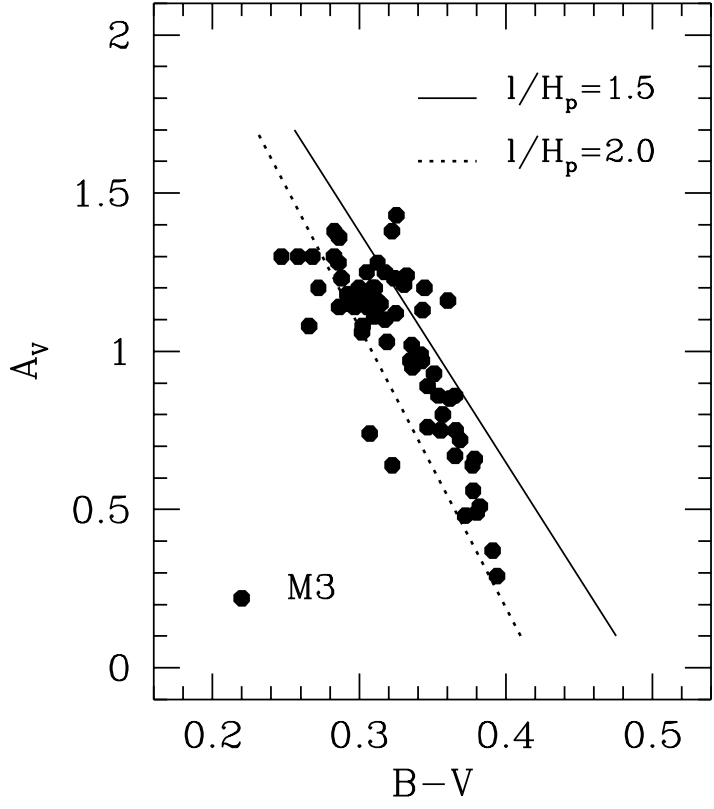
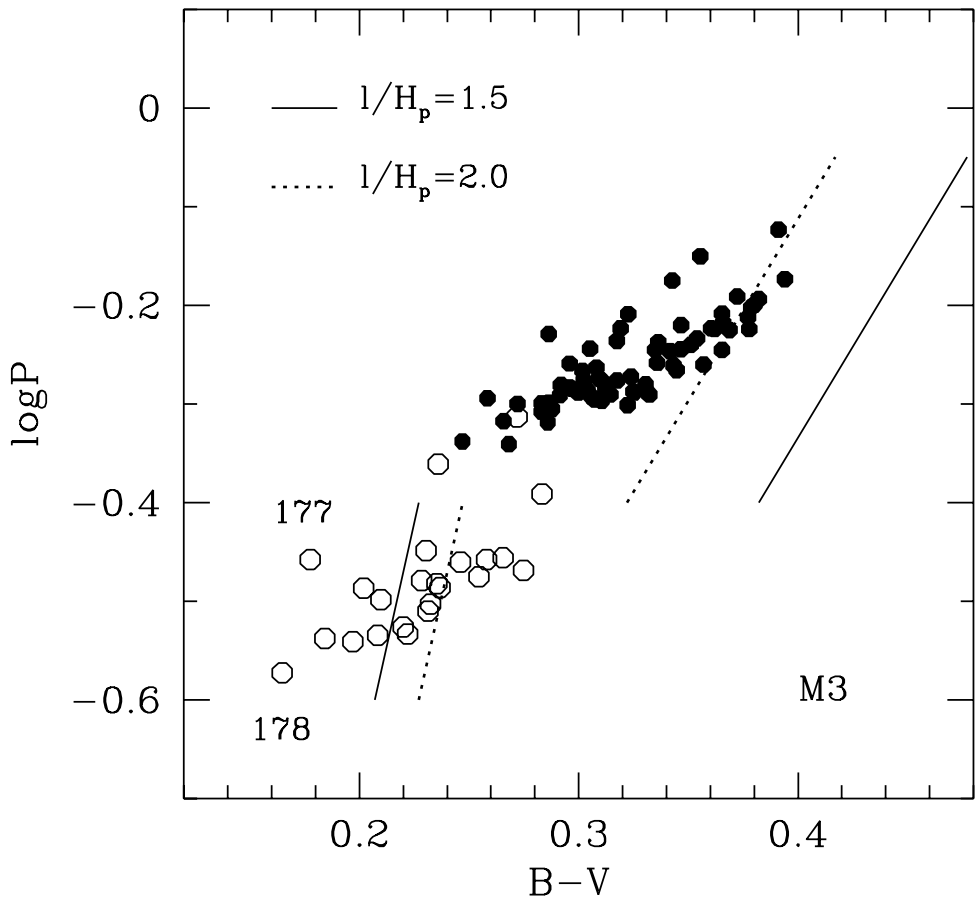
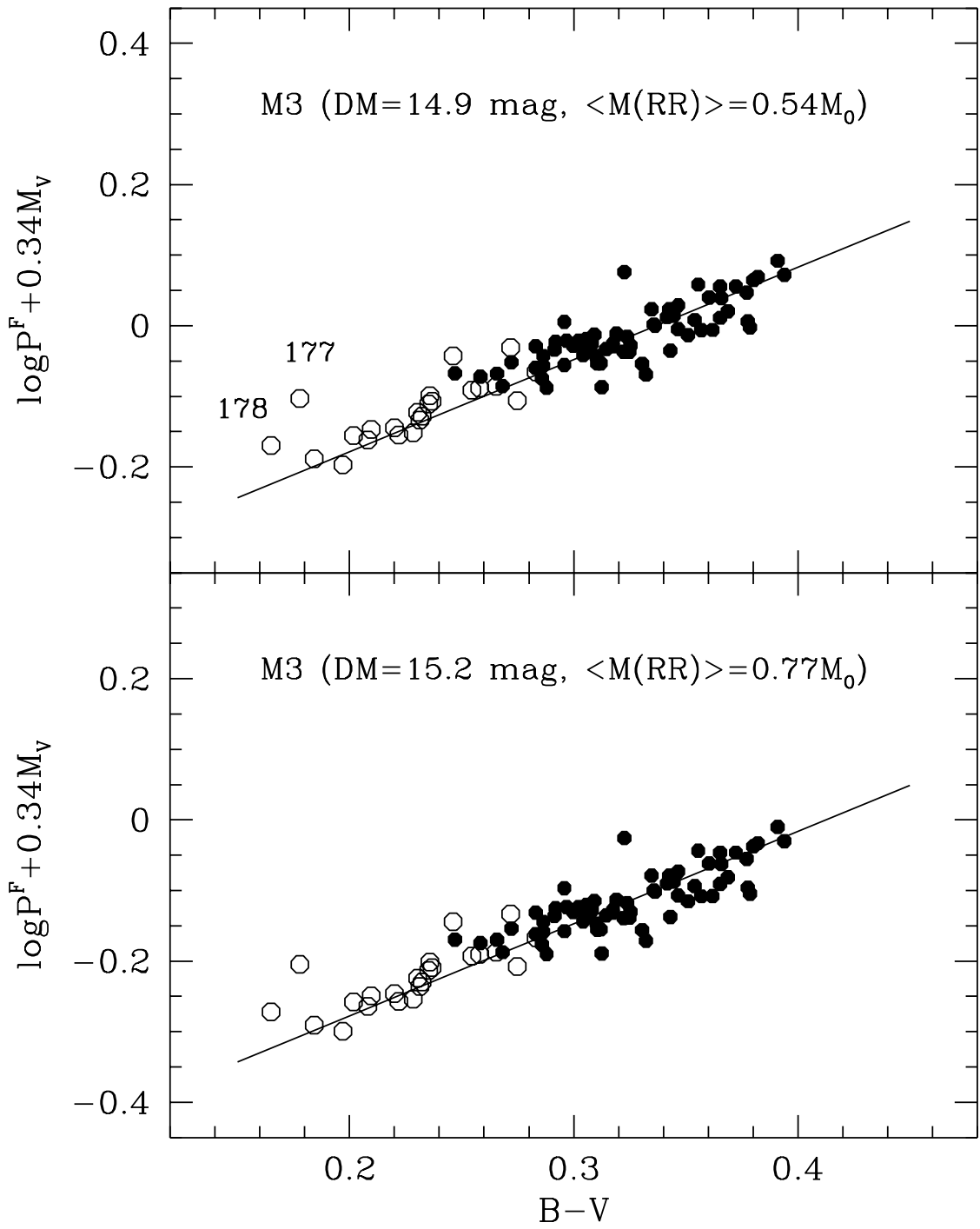
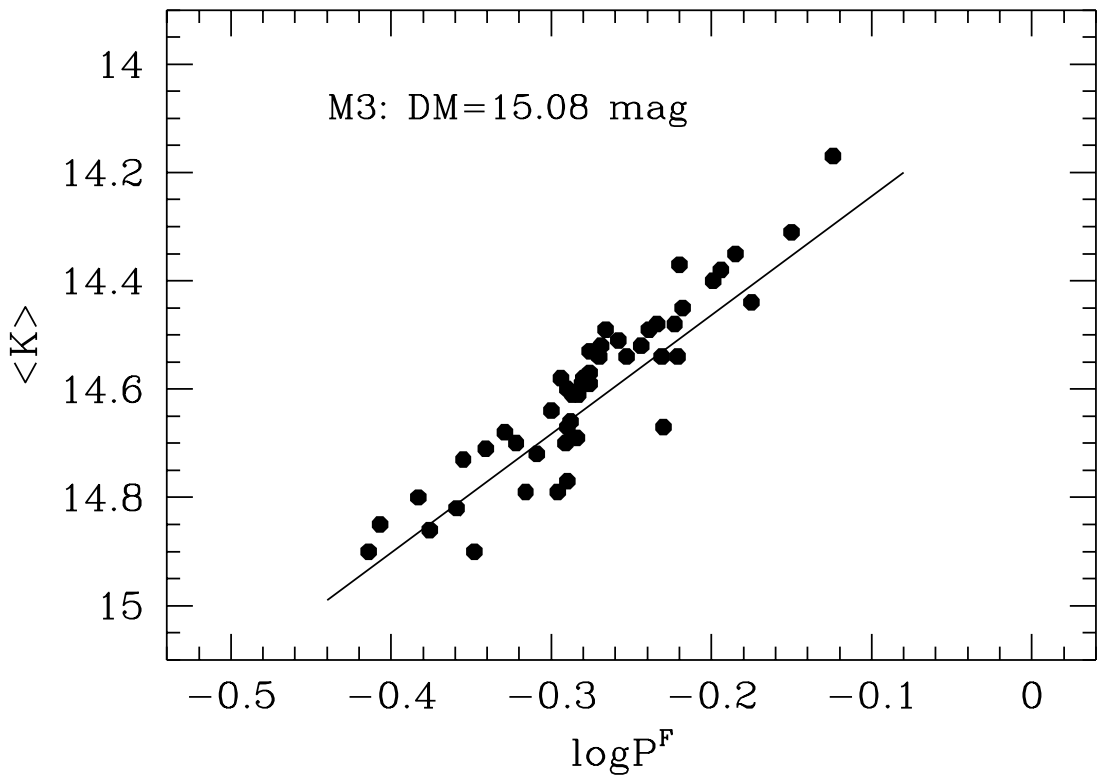
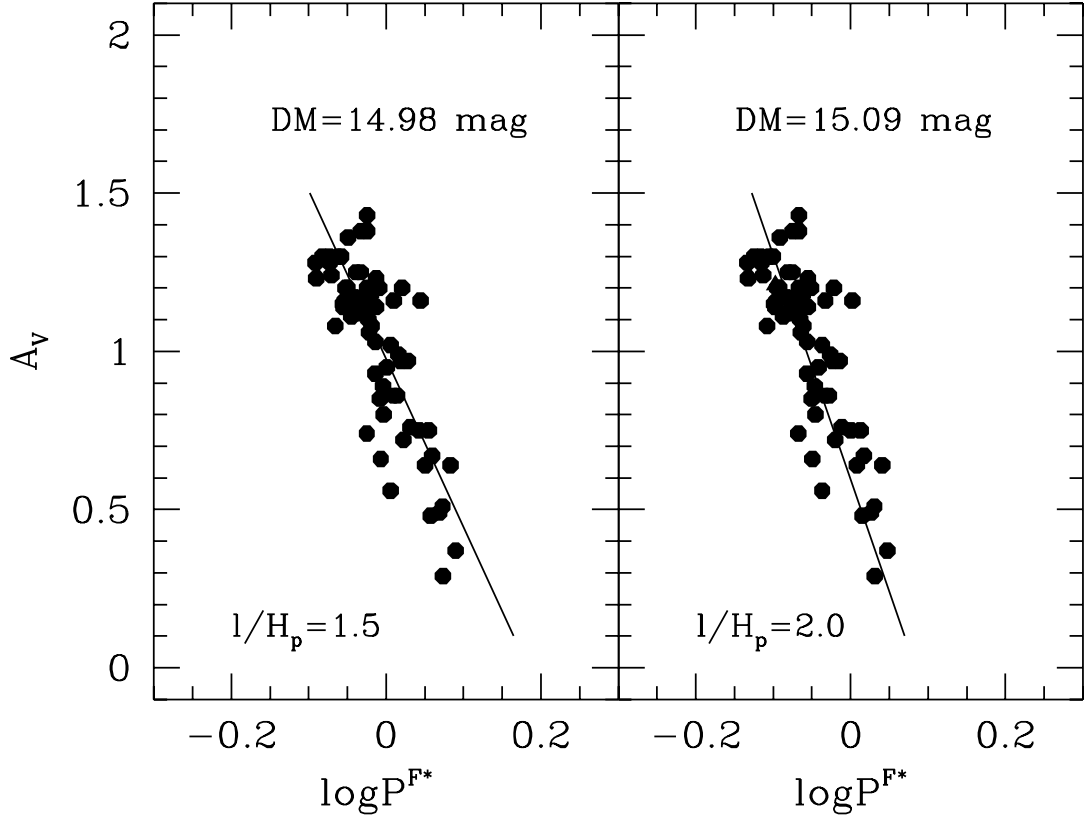
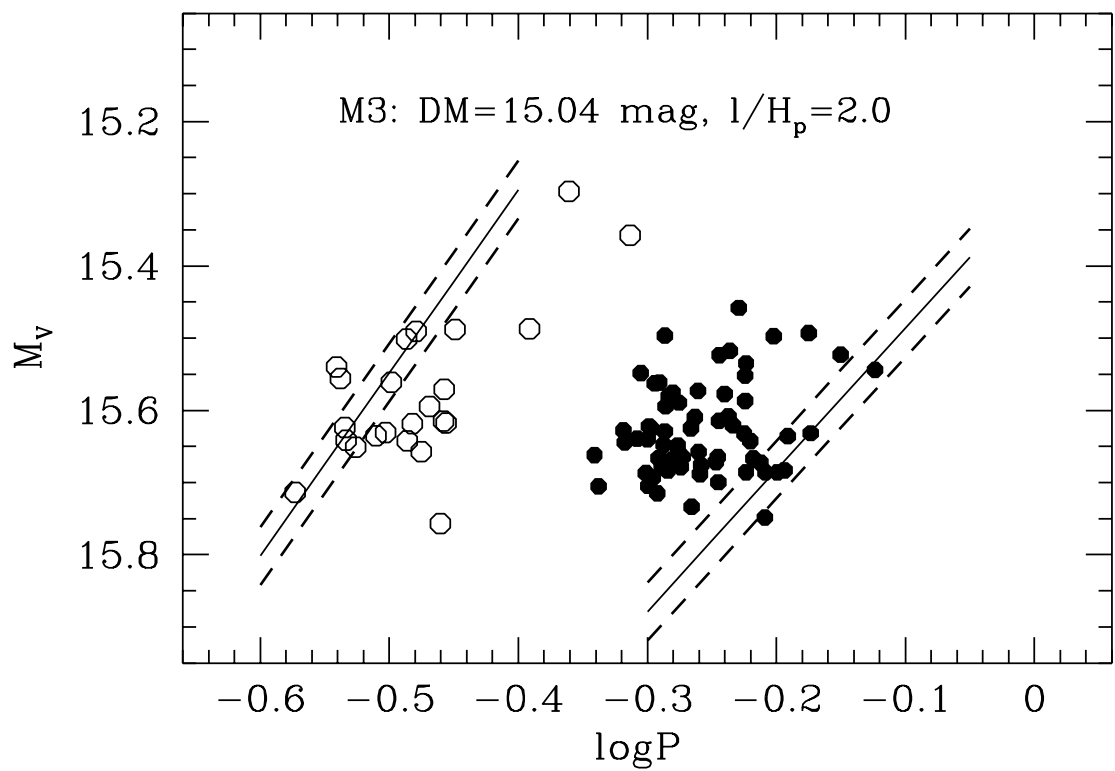
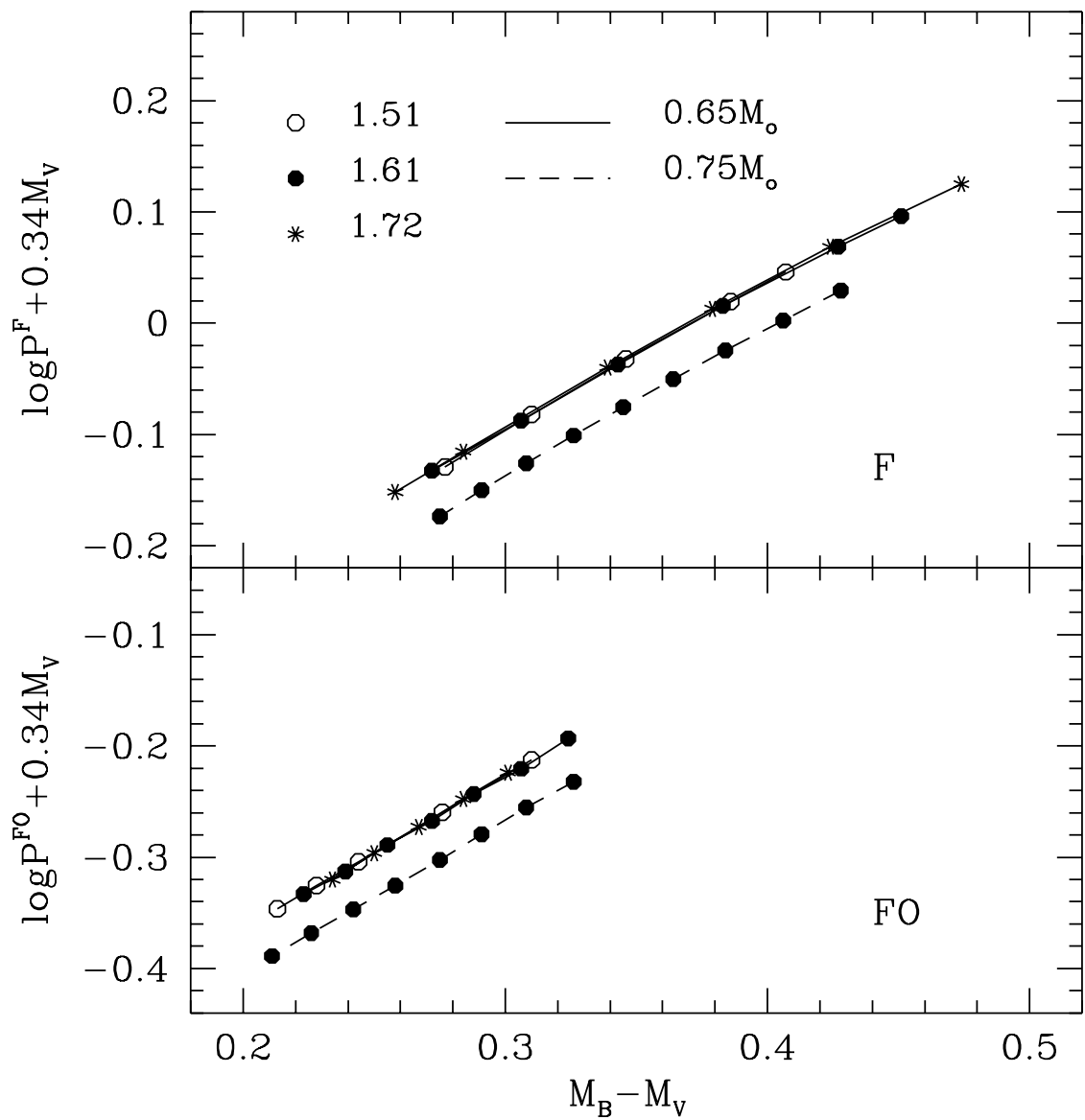
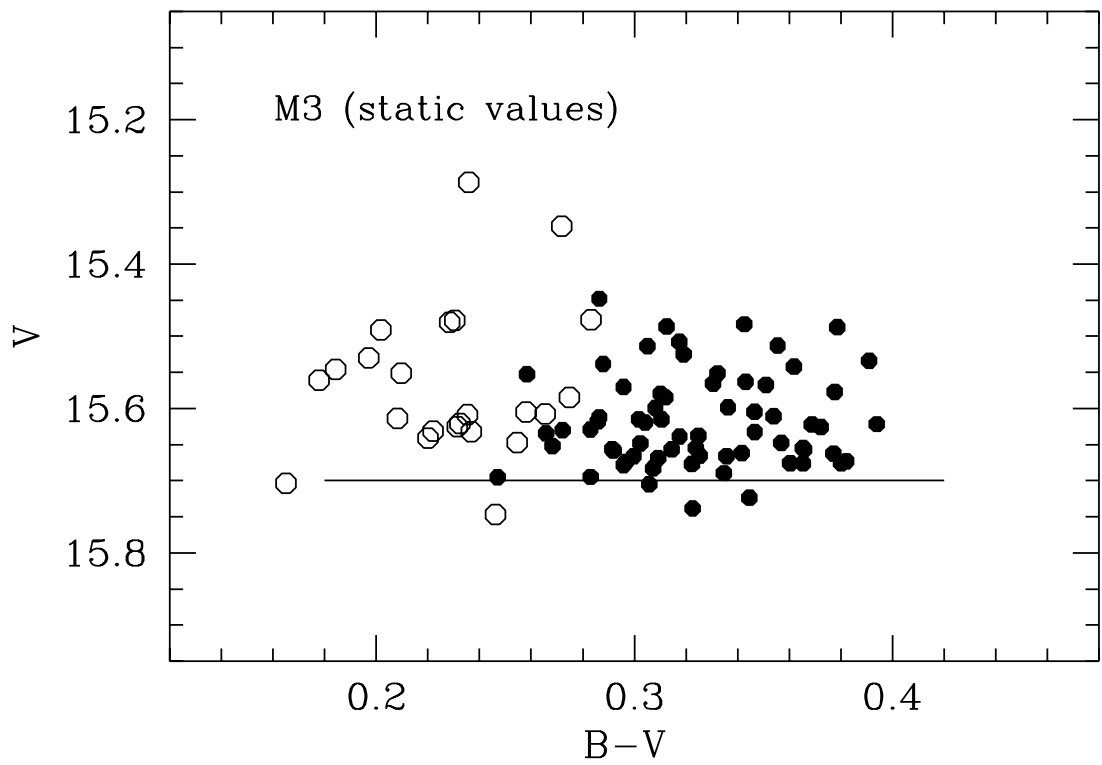
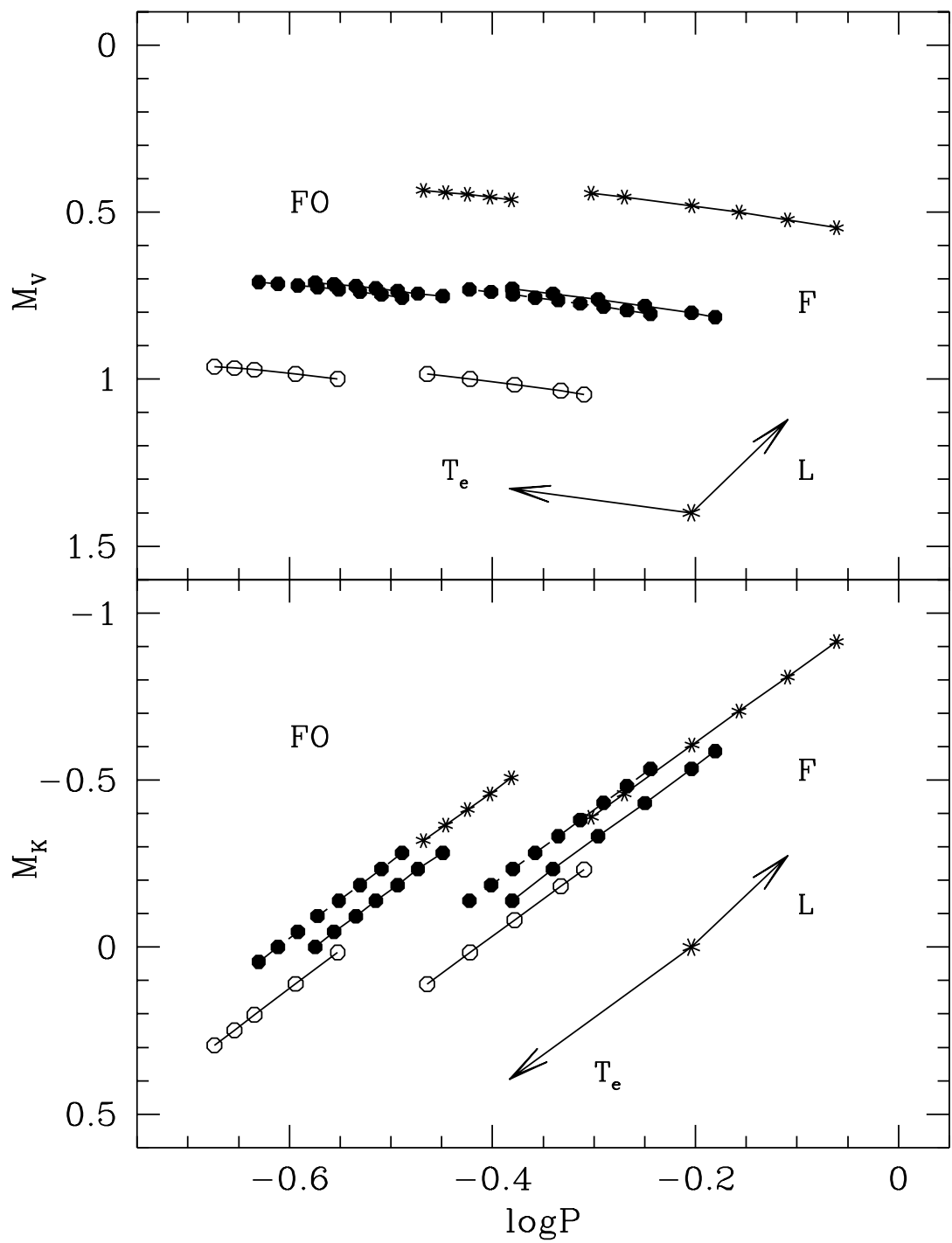
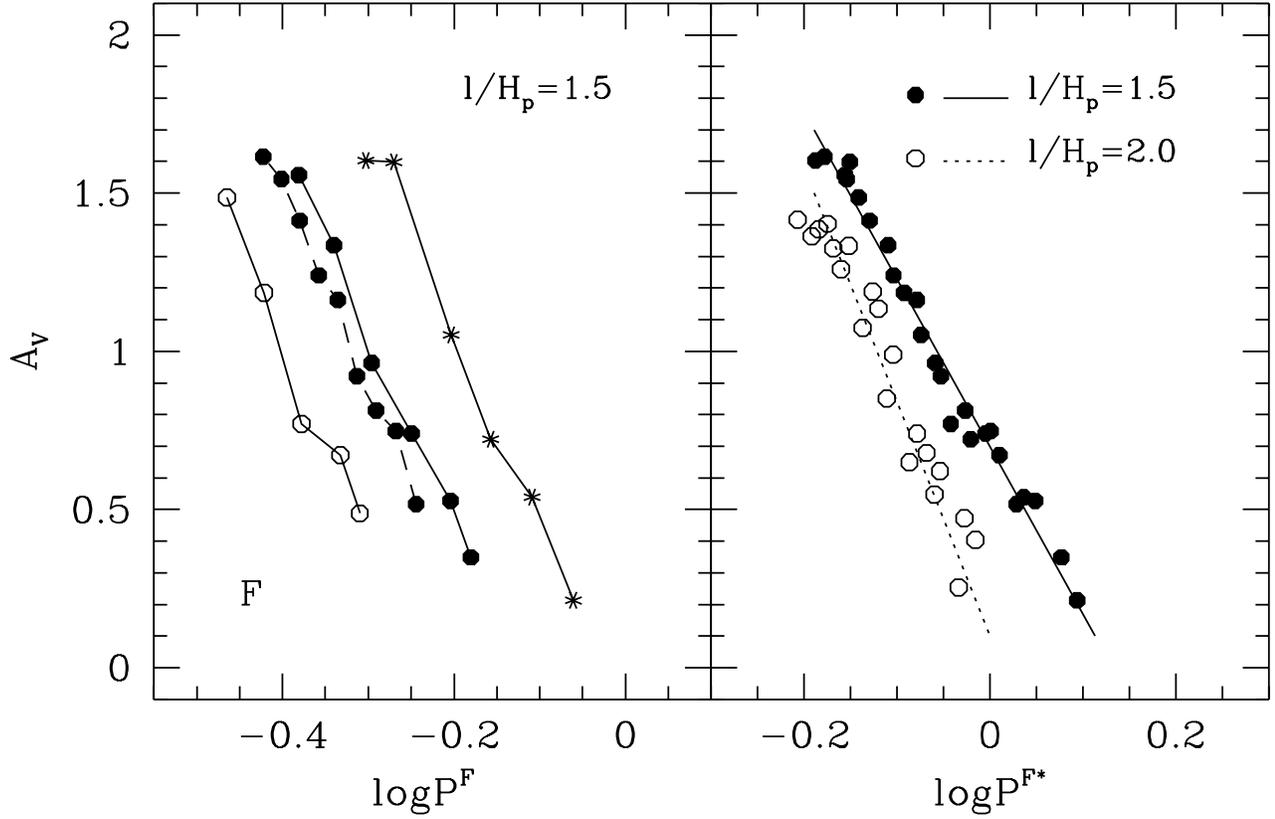
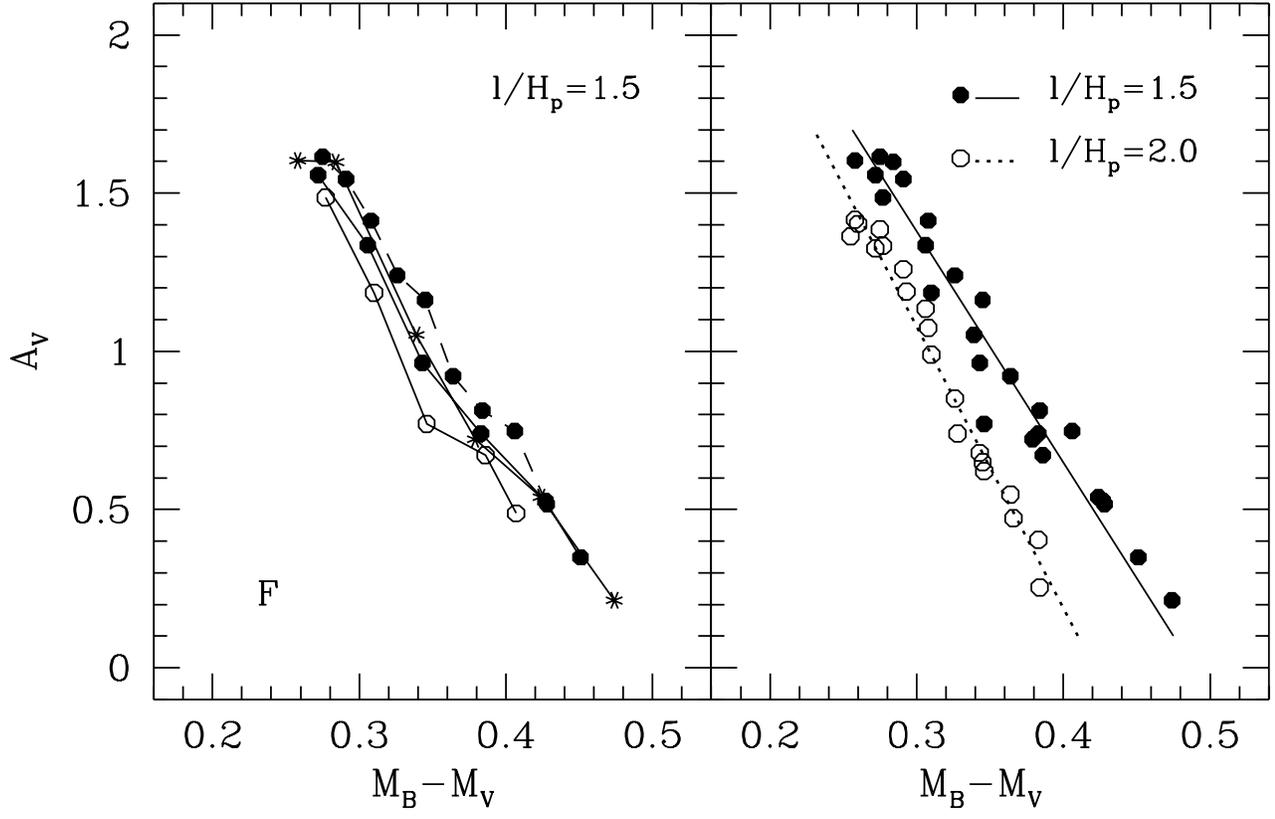
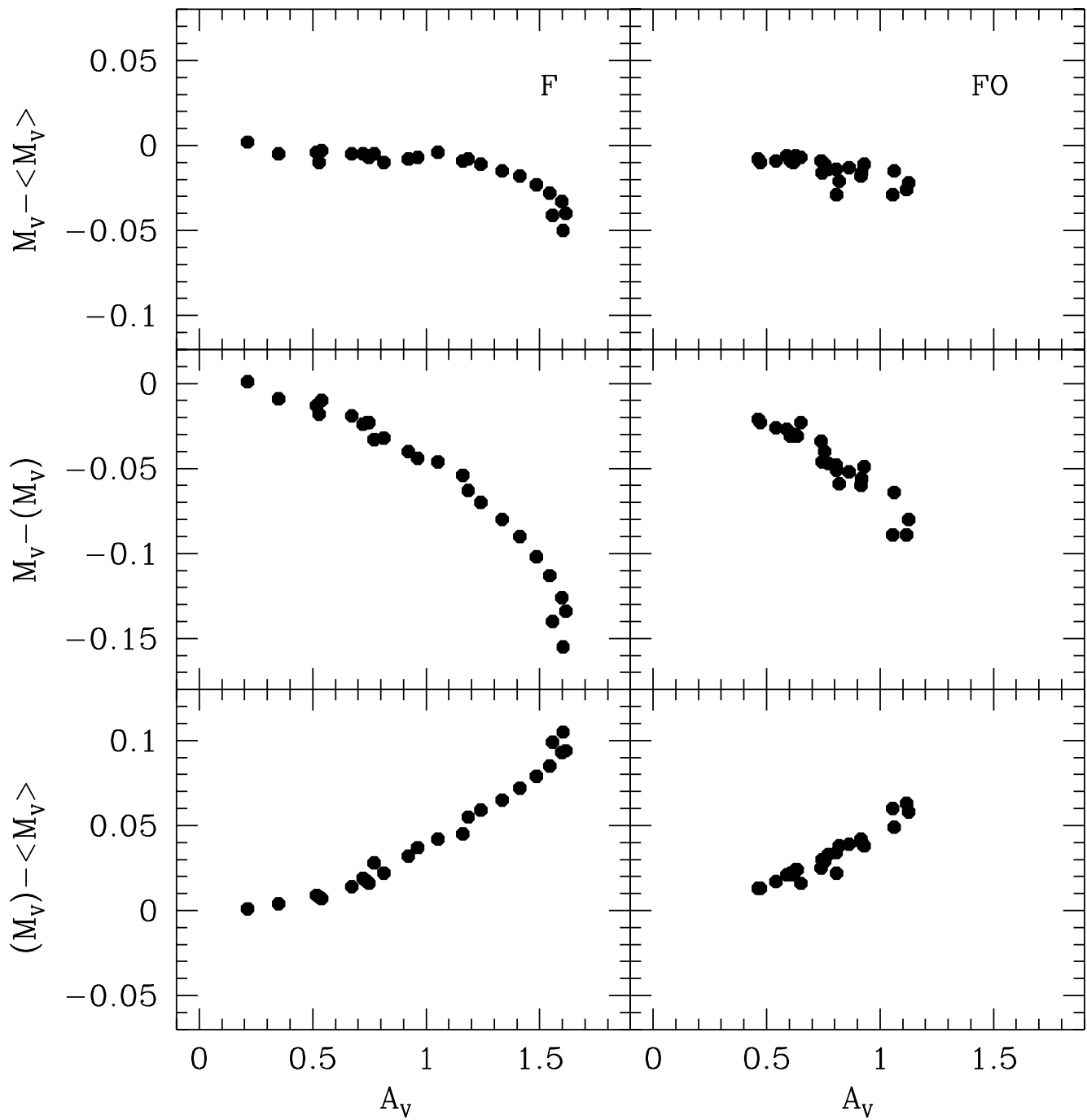
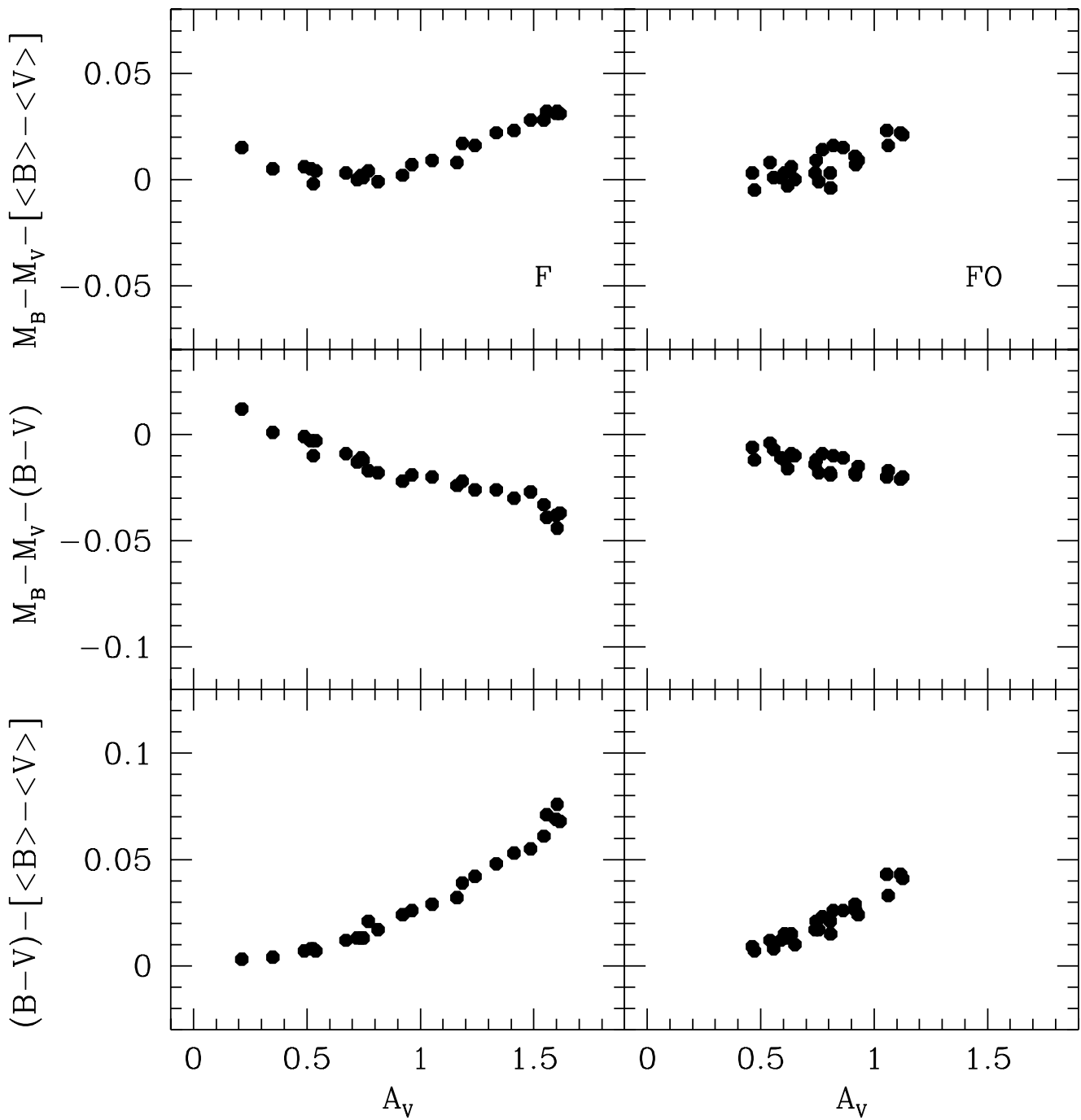
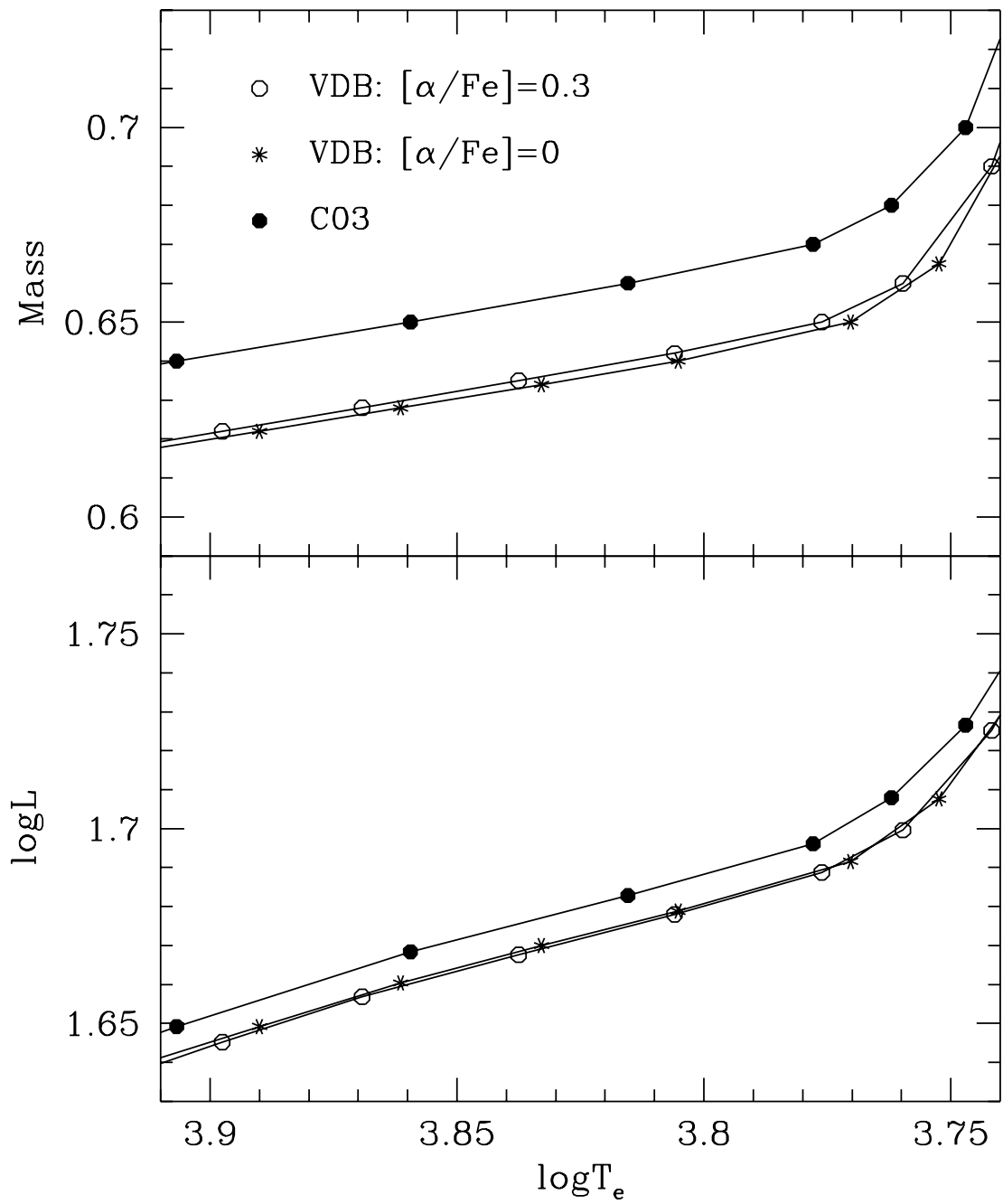
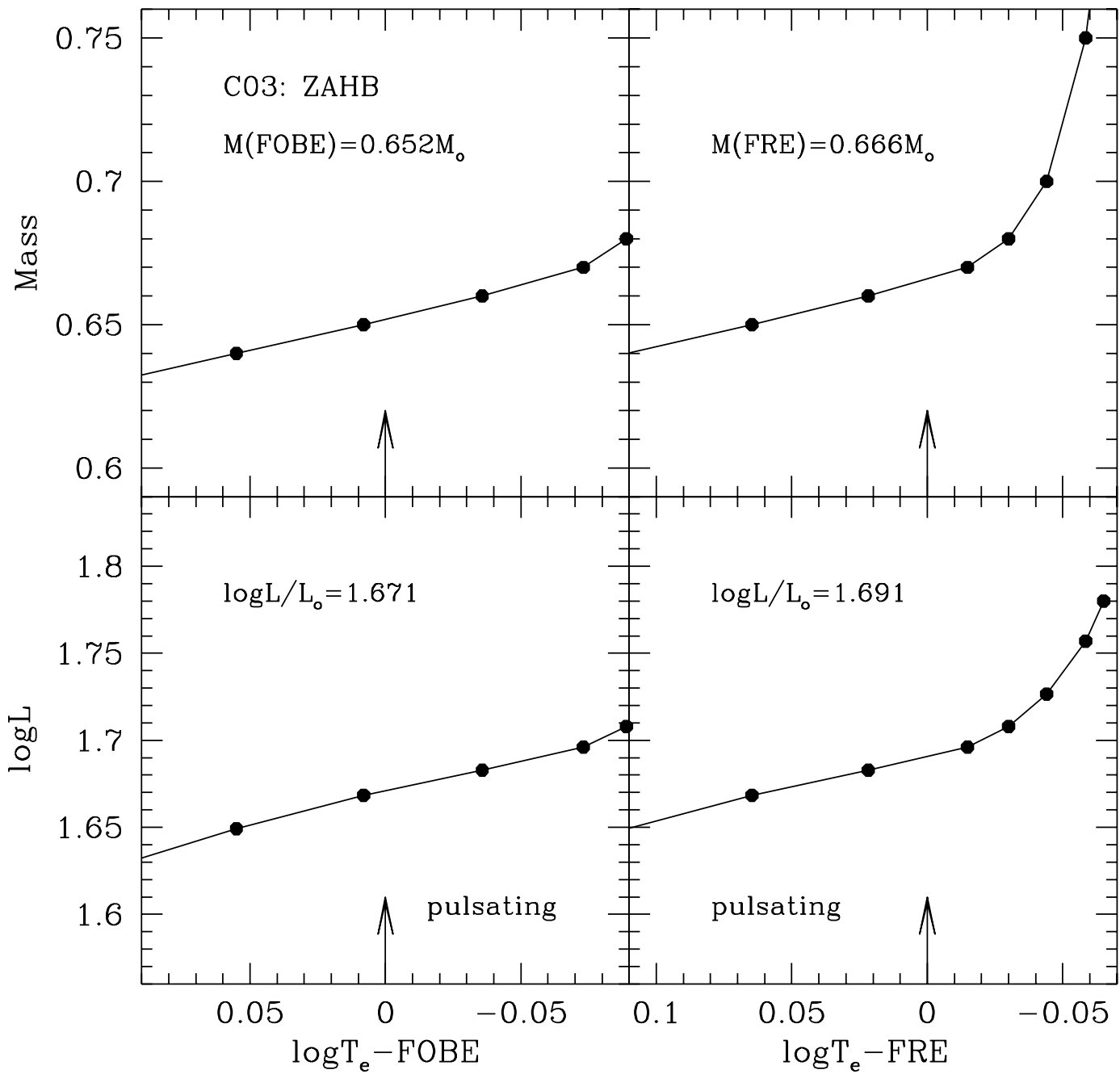
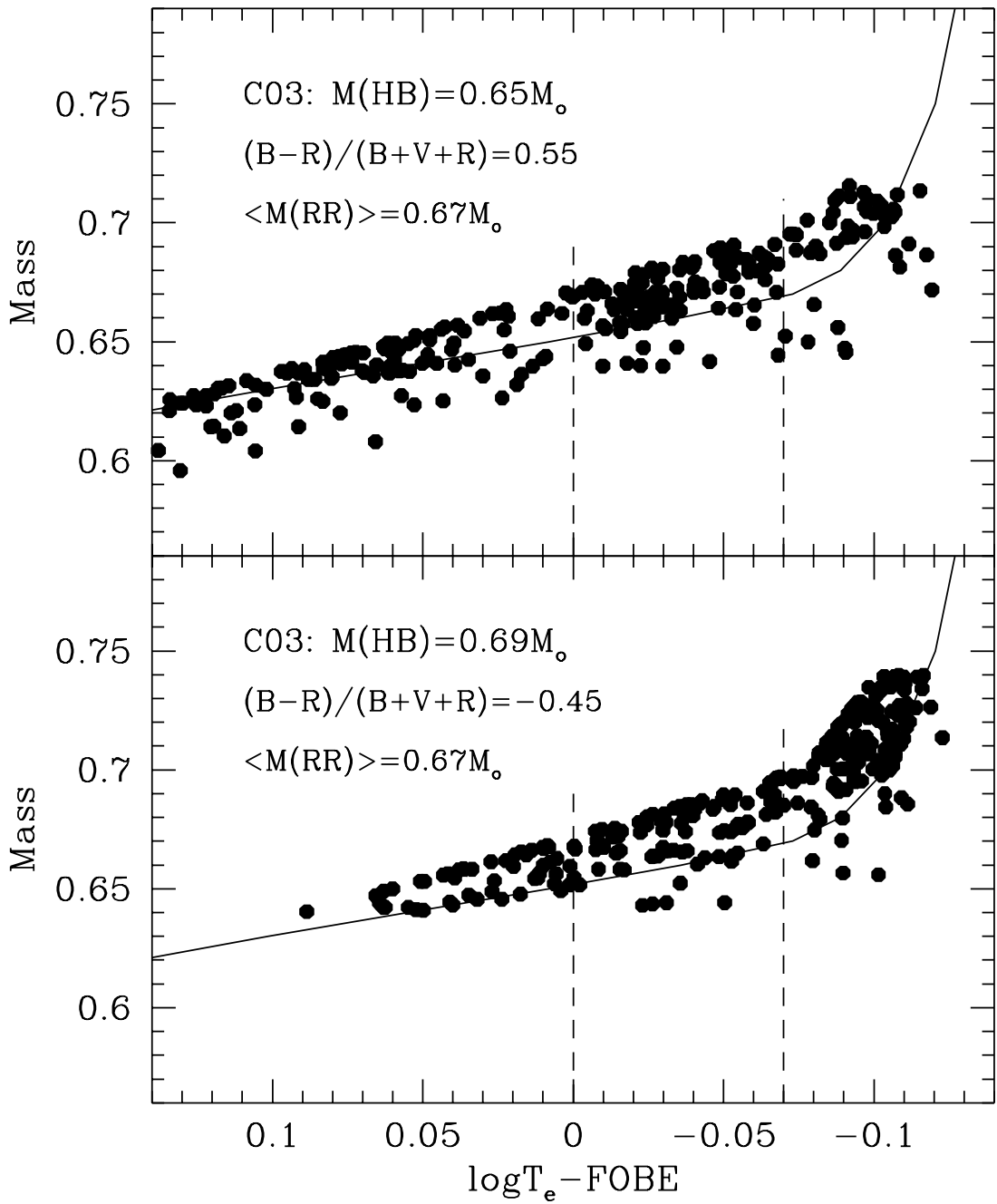
\end{document}